%% file: main.tex
\documentclass[10pt,journal,letterpaper]{IEEEtran}

\usepackage[utf8]{inputenc}
\usepackage{cite}
\usepackage{soul, color}
\usepackage{graphicx}
\usepackage{subcaption}
\usepackage{float}
\usepackage[colorlinks=true, urlcolor=blue, linkcolor=red]{hyperref}
\usepackage{booktabs}
\usepackage{adjustbox}
\usepackage{amsmath}
\usepackage{array}
\usepackage{tabularx}
\usepackage{multirow}
\usepackage{fancyhdr}
\usepackage{rotating}
\usepackage{stfloats}
\usepackage[dvipsnames]{xcolor}
\usepackage{siunitx}
\DeclareSIUnit\bit{bit}
\usepackage{amsfonts}
\usepackage{microtype}
\usepackage{mathptmx}   
\usepackage{xspace}
\newcommand{\lchem}{\emph{xCHEM}\xspace}

\input{Figures/tikz_preamble}

\usepackage[acronyms,nonumberlist,nopostdot,nomain,nogroupskip,acronymlists={hidden}]{glossaries}
\newglossary[algh]{hidden}{acrh}{acnh}{Hidden Acronyms}
\glsdisablehyper
\input{acronym}

\iftrue
\usepackage{tikzpagenodes,etoolbox}
\usetikzlibrary{calc}
\usepackage[contents={}]{background}
\AddEverypageHook{%
\ifnumequal{\thepage}{1}{%
    \tikz[remember picture,overlay]{%
        \node[
            draw,
            minimum width=0.9\textwidth,
            text width=0.9\textwidth,
            align=center,
            font=\footnotesize
        ]
        at ($(current page header area) - (0,-1pt)$)
        {%
            This work has been submitted to the IEEE for possible publication.
        };
    }%
}{}
}
\fi

\begin{document}

\title{\lchem: SDR-Based Multi-Link Channel Emulator for
Real-Time Hardware-in-the-Loop Digital-Twin}

\author{Ali Saeizadeh, Gabriele Gemmi, Tamerlan Aghayev, Michele Polese, Pedram Johari, and Tommaso Melodia%
\thanks{This work was supported in part by the National Telecommunications and Information Administration (NTIA)'s Public Wireless Supply Chain Innovation Fund (PWSCIF) under Award No. 25-60-IF011. (Corresponding author: Tommaso Melodia.)}%
\thanks{The authors are with the Institute for Intelligent Networked Systems, Northeastern University, Boston, MA 02115, USA (e-mail: saeizadeh.a@northeastern.edu; g.gemmi@northeastern.edu; aghayev.t@northeastern.edu; m.polese@northeastern.edu; p.johari@northeastern.edu; t.melodia@northeastern.edu).}%
}
\maketitle

\begin{abstract}
This paper presents \lchem, a radio-frequency channel emulator that inserts programmable channels between physical radios using a single NI USRP X4xx software-defined radio.
External channel models stream impulse responses to its field-programmable gate array without firmware changes or interruption of signal processing.
A streaming datapath implements six independently configurable directional channels across four radio interfaces.
Each channel uses a sparse finite impulse response filter whose memory-backed delays decouple delay span from multiplier resources.
The real and complex variants support 32 and 16 nonzero taps, respectively, with a maximum delay of \SI{16666.62}{\nano\second} and \SI{4.069}{\nano\second} resolution.
Coefficient scaling and arithmetic right shifts control attenuation.
Hardware measurements show root-mean-square errors of \SI{9.5}{\pico\second} in delay over the full span and \SI{0.11}{\decibel} in attenuation.
Multipath delays are recovered within one sample. An FPGA coefficient engine produces tested Doppler shifts from $\pm\SI{1}{\hertz}$ to $\pm\SI{1}{\mega\hertz}$ with mean frequency errors below \SI{0.032}{\hertz} relative to quantized commands after carrier-offset calibration; at \SI{200}{\hertz}, it reduces the measured near-carrier spur by \SI{53.3}{\decibel} compared with \SI{1}{\kilo\hertz} host updates.
\end{abstract}



\section{Introduction}
\label{sec:intro}

As large language models and agentic \gls{ai} accelerate the generation of wireless software and control policies, verifying their behavior on real hardware becomes increasingly important~\cite{aghayev2026genesis}.
For 5G-Advanced and 6G networks, this requires realistic, repeatable propagation conditions: simulation alone cannot establish hardware behavior, while \gls{ota} experiments are difficult to reproduce across sites and time~\cite{polese2024colosseum}.
\gls{hil} RF channel emulation provides programmable delay, attenuation, multipath, and Doppler between physical radios over cabled links, enabling repeatable testing of implementations and policies before deployment.

The state of the art in \gls{hil} channel emulators today falls into three disjoint camps. 
\emph{Commercial appliances}---Keysight PROPSIM~\cite{PropsimF64} and Spirent Vertex~\cite{spirentVertex}---support programmable propagation effects and multi-channel testing through vendor-provided hardware and control interfaces. The cited sources do not establish public access to their channel-processing \gls{rtl}.
\emph{Large testbeds}---most prominently the Colosseum wireless network emulator~\cite{polese2024colosseum} 
provide large-scale programmable \gls{hil} environments but require dedicated \gls{fpga} farms and centralized infrastructure.
\emph{Bespoke hardware emulators} address specific bandwidths, fading models, and applications~\cite{olmos1999design,eslami2009design,hofer2018real,mao2024rfemu,fei2026ris}; Section~\ref{sec:related_work} reviews their relationship to an open, multi-port benchtop platform.
A recent open-source effort, \gls{oal}~\cite{deshpande2024openairlink}, brings the methodology to off-the-shelf \glspl{usrp} and is the closest precedent to the work presented here; however, its dense \gls{fir} channel block consumes one \gls{dsp} slice per delay step and supports only real-valued coefficients, restricting it to short propagation spans (tens of taps, sub-microsecond delay) on a single link.

In parallel, advances in physics-based and learned propagation models are making site-specific wireless scenarios increasingly accessible.
Ray-tracing tools such as Wireless InSite~\cite{remcomWI} and Sionna~RT~\cite{sionna} generate detailed \glspl{cir}, while AI-based approaches such as AIRMap~\cite{saeizadeh_airmap_2026} estimate path-gain radio maps with millisecond-scale inference.
These advances motivate a complementary capability: an open, live interface through which channel models can drive physical radios on an accessible benchtop platform.

We present \lchem, a channel emulator that combines multi-channel operation, microsecond-scale delay spans, per-path Doppler, and live \gls{cir} updates on a single benchtop SDR, built on commodity hardware and a standard \gls{rfnoc} toolchain, with channel-emulation blocks that have no dependencies on proprietary IP cores.
Its model-independent control interface connects external channel models to physical radios, enabling reproducible evaluation of wireless implementations and control policies.

\noindent\textbf{Contributions.} This paper makes three contributions:
\begin{itemize}

\item \emph{An integrated multi-channel SDR architecture.}
We design and implement six independently configurable directional channels on a single USRP~X410, each incorporating a sparse, \gls{bram}-backed complex \gls{fir} filter. The filter architecture decouples the programmable delay span from the available multiplier budget, enabling \lchem to support delays of up to \SI{16666.624}{\nano\second} with \SI{4.069}{\nano\second} resolution. An FPGA engine in each complex filter synthesizes up to four Doppler components per tap, with sample-indexed phase evolution, interpolated coefficients, and timestamped atomic parameter commits.

\item \emph{Live integration of external channel models.}
We develop a streaming control interface that updates channel parameters without interrupting the sample stream.
The interface accepts tapped-delay \glspl{cir} within the hardware's tap-count, delay, and update-rate limits, allowing external models to drive the emulator without firmware changes.
Reference Sionna~RT and 3GPP channel-model front-ends connect scenario generation to physical RF emulation.


\item \emph{Channel validation and experimental 5G platform.}
We characterize delay, attenuation, multipath, multi-channel operation, Doppler, and four-component fading through hardware experiments.
Measurements yield a delay RMSE of \SI{9.5}{\pico\second} over the full \SI{16666.624}{\nano\second} span and an attenuation RMSE of \SI{0.11}{\decibel} over \SIrange{6}{42}{\decibel}.
We also present a RAN verification setup combining a commercial \gls{ru}, an \gls{oai} \gls{cu}/\gls{du} stack, and \gls{cots} \glspl{ue}.
Together, the channel validation and integration details provide a practical foundation for researchers to establish 5G experiments under programmable channel conditions.

\end{itemize}

\section{Related Work}
\label{sec:related_work}

Table~\ref{tab:related_work} compares representative hardware emulators; below, we position \lchem relative to these architectures and shared experimentation platforms.

\begin{table*}[!t]
  \centering
  \caption{Comparison of representative RF channel emulation platforms.}
  \label{tab:related_work}
  \scriptsize
  \setlength{\tabcolsep}{4pt}
  \renewcommand{\arraystretch}{1.12}
  \begin{tabular*}{\textwidth}{@{\extracolsep{\fill}}lllllll@{}}
    \toprule
    \textbf{Emulator} & \textbf{Platform} & \textbf{Configuration} & \textbf{BW (MHz)} & \textbf{Taps/paths} & \textbf{Delay span} & \textbf{Doppler} \\
    \midrule
    PROPSIM F64~\cite{PropsimF64} & Appliance & Massive MIMO & Up to 1200$^a$ & Up to 48$^a$ & 335\,$\mu$s; 1.5\,s opt.$^a$ & Yes \\
    Vertex~\cite{spirentVertex} & Appliance & Up to 256 links$^a$ & 40--1200$^a$ & Up to 24$^a$ & 1--2\,s$^a$ & Yes \\
    Eslami et al.~\cite{eslami2009design} & \gls{fpga} & MIMO & --- & --- & --- & --- \\
    Hofer et al.~\cite{hofer2018real} & SDR+host & Single link & 10 (evaluated) & --- & --- & Yes \\
    Colosseum/MCHEM~\cite{polese2024colosseum,colosseumFAQ} & FPGA cluster & $256\!\times\!256$ mesh & 80 & 4 nonzero, C & 5.12\,$\mu$s & CIR updates \\
    OpenAirLink~\cite{deshpande2024openairlink} & \gls{usrp} X310 & 2 directional links & --- & 41 dense, R & 0.167\,$\mu$s & No \\
    RF Emu II~\cite{mao2024rfemu} & ASIC & 4-node prototype & 518 & DPCM & --- & --- \\
    \midrule
    \textbf{\lchem} & \textbf{USRP X4xx} & \textbf{6 directional links} & \textbf{---}$^b$ & \textbf{32R / 16C} & \textbf{16.66\,$\mu$s} & \textbf{FPGA engine} \\
    \bottomrule
  \end{tabular*}
  \par\vspace{2pt}
  \begin{minipage}{\textwidth}
    \scriptsize
    R/C: real/complex coefficients; DPCM: direct-path computational model; ---: comparable specification not established from the cited source.
    $^a$Commercial specifications depend on configuration and need not be simultaneous. Path counts are per fading channel (PROPSIM) or digital channel (Vertex). Vertex supports up to 256 digital links at 40/100\,MHz and 64 at 200\,MHz; its maximum delay is 2\,s at 40\,MHz and 1\,s at bandwidths of at least 100\,MHz. PROPSIM specifies a maximum delay of 335\,$\mu$s without extended-delay options; options F8800A-ED2 and F8800A-ED1 extend this to 100\,ms and 1500\,ms, respectively, with one digital channel per TRX channel when either option is active.
    $^b$\lchem operates at 245.76\,MS/s; sample rate is not a measured usable RF bandwidth. Real/complex variants are listed in that order.
  \end{minipage}
    \vspace{-.3cm}
\end{table*}

\subsection{FPGA and SDR Channel Emulators}

Early hardware channel emulators~\cite{olmos1999design,dassatti2005high,dung2006implementation} established programmable tapped-delay architectures that combine weighted, delayed copies of an input signal.
Later work extended this approach to non-stationary \gls{gbsm} channels~\cite{zhu2018novel}, geometry-driven \gls{cir} updates~\cite{hofer2018real,ghiaasi2018real,ghiaasi2016realtime}, and applications including vehicular communications~\cite{fernandez2009fpga,ruiz2019sdr}, \gls{mimo}~\cite{eslami2009design,picol2008mimo}, and EMC testing~\cite{patnaik2014implementation}.

Prior work also addresses FPGA-based network emulation~\cite{borries2009fpga}, hardware-in-the-loop SDR channels~\cite{matai2012designing}, and accuracy, range, and resource-scaling tradeoffs~\cite{huang2013towards,buscemi2013range}.
Other efforts emphasize scale, bandwidth, or specialized processing.
Chaudhari and Braun~\cite{ChaudhariB18} develop a scalable \gls{fpga} architecture for large channel-emulation systems, now used in Colosseum, while Matsumura~\textit{et~al.}~\cite{matsumura2022large} combine physical \glspl{sdr} with virtualized radios.
Mao~\textit{et~al.}~\cite{mao2024rfemu} achieve \SI{518}{\mega\hertz} per-channel bandwidth using a near-memory ASIC accelerator, and Fei~\textit{et~al.}~\cite{fei2026ris} dynamically allocate processing resources for RIS-aided channel emulation.

OpenAirLink (OAL)~\cite{deshpande2024openairlink} is the closest open-source precedent, implementing a dense, 41-tap real-coefficient \gls{fir} as an \gls{rfnoc} block on a \gls{usrp}.
\lchem builds on this approach, including its programmable right-shift attenuation, with three architectural extensions.
First, a sparse \gls{fir} with a \gls{bram}-backed circular delay buffer supports up to 16 nonzero complex taps over a maximum delay of \SI{16666.624}{\nano\second}, approximately 100 times OAL's \SI{167}{\nano\second}.
Multiplier cost scales with the supported nonzero tap count, while \gls{bram} capacity determines the addressable delay span.
Second, complex coefficients enable per-path Doppler through FPGA coefficient synthesis.
Third, a cascaded split--combine datapath integrates six independently configurable directional channels across the four RF interfaces of a single X410.

\subsection{Commercial RF Channel Emulators}

Commercial RF channel emulators are direct alternatives for repeatable testing of physical radios under controlled propagation conditions.
Keysight PROPSIM F64~\cite{PropsimF64} supports multi-channel and massive-\gls{mimo} testing with standardized channel models, while Spirent Vertex~\cite{spirentVertex} reproduces propagation effects including multipath, delay, and Doppler for wireless device evaluation.
These systems establish that programmable RF propagation and multi-channel operation are mature capabilities; neither capability alone distinguishes \lchem from commercial alternatives.

\lchem instead prioritizes a modifiable FPGA datapath, host software, and control protocol on a single commodity \gls{sdr} for external-model integration and automated testing~\cite{aghayev2026genesis}. It does not claim to match commercial instruments' full bandwidth, channel scale, or standardized test coverage.

\subsection{Large-Scale Wireless Experimentation Platforms}

Shared wireless platforms support experiments that extend beyond individual channel-emulation architectures.
Colosseum~\cite{chaudhari2018colosseum,polese2024colosseum} combines 256 \glspl{sdr} with a centralized MCHEM core to apply programmable propagation conditions across a large network of physical radios.
Its use as a digital twin for Open RAN~\cite{polese2024colosseum} illustrates the role of controlled RF emulation in evaluating network behavior.
Scenario-generation and integration tools further connect this infrastructure to external environments: CaST~\cite{villa2022cast} generates channel-emulation scenarios, ColosSUMO~\cite{gemmi_colossumo_2024} integrates vehicular mobility, and TwiNet~\cite{robinson_twinet_2024} links physical and digital wireless environments.

Field testbeds address a complementary experimental need by exposing radios to physical deployment environments.
POWDER~\cite{breen2021powder}, COSMOS~\cite{raychaudhuri2020challenge}, AERPAW~\cite{sichitiu2020aerpaw,panicker2021aerpaw}, and ARA~\cite{zhang2021ara} support metropolitan, urban, aerial, and rural wireless research, respectively.
Their field deployments capture propagation and operational conditions that are difficult to reproduce in a laboratory, whereas controlled channel emulation allows selected propagation parameters to be varied and replayed across experiments.
\lchem targets local development and repeatable hardware testing by integrating configurable RF channels into a single benchtop \gls{sdr}.

\section{System Architecture}
\label{sec:architecture}

\begin{figure}[t]
    \centering
    \input{Figures/lchem-3ch}
    \vspace{-.5cm}
    \caption{\lchem architecture: streamed channel profiles reach the embedded CPU (host), which writes FPGA control registers. Dashed arrows denote control; solid arrows show the three downlink and three uplink signal paths.}
    \label{fig:arch}
    \vspace{-.3cm}
\end{figure}
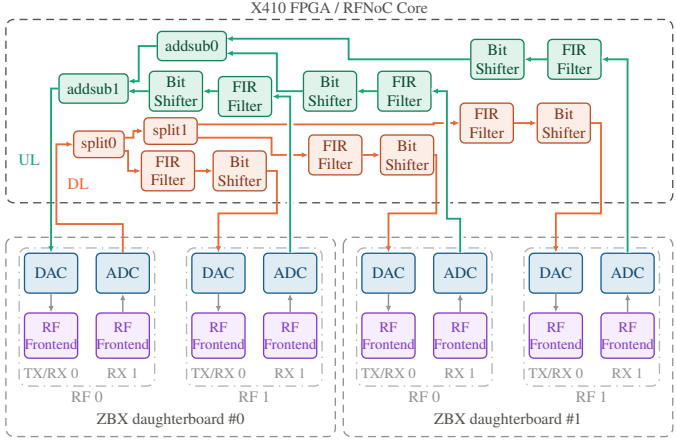

The \lchem configuration in Fig.~\ref{fig:arch} implements six directional channels across four RF interfaces on a single \gls{usrp} X410.
This one-to-three topology serves the experimental requirements of this study; other node configurations can be realized by adapting the \gls{rfnoc} signal graph within the available RF-interface and FPGA-resource limits.
The FPGA processes RF samples, while the emulator application on the X410's embedded CPU updates channel parameters through UHD/RFNoC register writes.
Throughout this paper, \emph{host} denotes this CPU and its control software. An external scenario computer can stream channel profiles to the application over ZeroMQ; RF samples remain in the FPGA datapath.

\subsection{RF Front-End}

The \gls{usrp} X410 integrates four transmit and four receive channels with an AMD Zynq UltraScale+ ZU28DR RFSoC, including 12-bit \glspl{adc} and 14-bit \glspl{dac}.
It supports an RF frequency range of \SI{1}{\mega\hertz} to \SI{7.2}{\giga\hertz}, up to \SI{400}{\mega\hertz} instantaneous bandwidth per channel, and IQ sample rates up to \SI{500}{\mega Sa\per\second}, depending on the FPGA image and clock configuration~\cite{x410HardwareManual}.
The \lchem images use the \SI{200}{\mega\hertz} RF configuration and operate at an IQ sample rate of $f_s = 245.76$\,MS/s, corresponding to a sample interval of \SI{4.069}{\nano\second}.

Received RF signals undergo frequency conversion, digitization, and digital downconversion to complex baseband.
The FPGA processes these samples as 16-bit I and 16-bit Q components, applying the emulated channel between the input $\hat{X}[n]$ and output $\hat{Y}[n]$ streams.
Digital upconversion, digital-to-analog conversion, and RF frequency conversion then produce the transmitted signal.

\subsection{RFNoC Signal Graph}
\label{sec:rfnoc_signal_graph}
\label{sec:variants}

As shown in Fig.~\ref{fig:arch}, the \gls{rfnoc} framework~\cite{braun_rfnoc_2016} connects the radio interfaces to streaming FPGA processing blocks.
In the downlink, two cascaded splitters distribute the gNB signal to three branches, each comprising a sparse \gls{fir} filter and a right-shift attenuator before transmission to a UE.
In the uplink, the three UE signals are independently filtered and attenuated, then combined by two cascaded adders for transmission to the gNB.
The unequal splitter depths introduce a fixed inter-channel delay offset, which is accounted for during delay validation (Section~\ref{sec:verification}).

Two FPGA images implement this graph with either 32 real-valued or 16 complex-valued taps per channel.
Both retain the same routing and attenuation stages and use \gls{bram}-backed delay buffers; the complex-coefficient variant additionally supports per-path Doppler through FPGA coefficient synthesis.
The following subsections describe the filter arithmetic and programmable delay design.

\subsection{Sparse FIR Architecture}
\label{sec:fir_design}

The channel filter separates the number of modeled paths from the length of the sample history needed to realize their delays.
A sparse tapped-delay structure allocates multiplication resources to each supported path, while independently addressed \gls{bram} buffers provide its delay.
Thus, extending the delay span increases memory demand without increasing the number of multipliers, provided that the supported tap count remains fixed.
For $N_\mathrm{tap}$ tap slots, the implemented channel is
\begin{equation}
  \hat{Y}[n] = \mathcal{Q}_{16}\!\left(2^{-15}
  \sum_{i=0}^{N_\mathrm{tap}-1} b_i[n]\hat{X}[n-d_i]\right),
  \label{eq:tdl}
\end{equation}
where $n$ indexes accepted input samples, $d_i$ is an integer sample delay, and $b_i[n]$ contains one signed 16-bit coefficient word in the real variant or a pair of such words in the complex variant.
The factor $2^{-15}$ accounts for the Q1.15 representation; $\mathcal{Q}_{16}$ rounds and saturates each output component to 16 bits.
Setting a coefficient to zero disables its contribution, but does not release the hardware allocated to that tap.

\subsubsection*{Coefficient arithmetic}
For real coefficients, two multipliers per tap scale I and Q independently.
Complex coefficients require four real multiplications and a registered cross-combination stage to form the real and imaginary products.
The corresponding multiplier counts are $2N_\mathrm{tap}$ and $4N_\mathrm{tap}$, independent of the delay-buffer depth.

The filter retains full-precision products through a balanced adder tree, using 32-bit products and $\log_2N_\mathrm{tap}$ accumulation guard bits.
The complex cross-combination adds one further guard bit.
Normalization occurs once after accumulation: the \gls{rtl} adds $2^{14}$, discards the 15 fractional bits, and saturates to the signed 16-bit range.
This implements round-to-nearest with ties toward positive infinity.
The rounding error is at most half an output LSB per component when saturation does not occur; this bound does not include coefficient quantization or clipping.

\subsubsection*{Pipeline timing}
The registered datapath comprises a memory read, a tap-data register, multiplication, an optional complex cross-combination, a balanced adder tree, and the output register.
Its stage count is
\begin{equation}
  L = 4+c_\mathrm{cplx}
      +\max\!\left(1,\lceil\log_2N_\mathrm{tap}\rceil\right),
  \label{eq:latency}
\end{equation}
where $c_\mathrm{cplx}=0$ for real coefficients and $1$ for complex coefficients.
Both deployed configurations have nine stages, independent of the programmed delays.
Because the datapath advances on accepted samples, this stage count must be distinguished from elapsed compute-clock cycles and from end-to-end RF latency, which also includes buffering, conversion, and routing.
The fixed system offsets are characterized separately during delay validation (Section~\ref{sec:verification}).

\subsection{Programmable Delay Design}
\label{sec:delay}

The delay subsystem supplies independently selected samples to the tap multipliers using replicated circular buffers.
Its memory organization determines the supported delay span while preserving parallel tap evaluation.

\subsubsection*{Independently addressed circular buffers}
Each tap has a separate copy of the I and Q sample histories, giving $2N_\mathrm{tap}$ memories of depth $D_\mathrm{max}$ and width 16 bits.
Replication supplies the independent read ports required to evaluate all tap delays in parallel.
All copies receive the same input sample at a common write address $w[n]$; tap $i$ reads from
\begin{equation}
  a_i[n] = (w[n]-d_i)\bmod D_\mathrm{max},
  \qquad 0\leq d_i<D_\mathrm{max}.
  \label{eq:delay_address}
\end{equation}
The depth is a synthesis-time power of two, so address truncation implements wraparound.
The write pointer advances only when the input valid/ready handshake accepts a sample, preserving sample-indexed delays across idle clock cycles.
For $d_i=0$, a registered forwarding path supplies the current sample instead of the stale value returned by a simultaneous read and write to the same address.
A further tap-data register separates this forwarding logic from the multipliers to ease timing closure.

Tap positions can be changed at runtime without shifting or rebuilding the stored history, and different taps may use the same delay.
The \gls{rtl} initializes the memories to zero, but a control reset resets the pointer and pipeline registers without clearing the memories; a full buffer of new samples replaces any retained history.
These buffers implement integer-sample delays; the datapath contains no fractional-delay interpolation.

\subsubsection*{Delay resolution and range}
The programmable delay grid follows the IQ sample rate rather than the FPGA compute-clock frequency:
\begin{equation}
  \Delta\tau = \frac{1}{f_s} \approx \SI{4.069}{\nano\second},
  \qquad f_s=245.76\,\mathrm{MS/s}.
  \label{eq:delay_res}
\end{equation}
Since the largest valid index is $D_\mathrm{max}-1$, the maximum additional path delay is
\begin{equation}
  \tau_\mathrm{max} = \frac{D_\mathrm{max}-1}{f_s}.
  \label{eq:delay_max}
\end{equation}
The deployed real and complex variants use depths of 4096 samples, yielding approximately \SI{16.66}{\micro\second} of programmable delay (Table~\ref{tab:resources}).
These spans correspond to approximately 5.0\,km of free-space propagation, excluding fixed hardware latency.
The \SI{4.069}{\nano\second} grid describes programmable tap spacing, not the ability of a bandwidth-limited measurement to resolve two paths.

The block-control API accepts sample indices and rejects delays outside the synthesized range.
The streaming interface also carries native sample indices: the sender converts physical delays to the nearest sample at $f_s=245.76$\,MS/s, or accepts indices directly.
The receiver applies these indices without rescaling; MT-012 clamps out-of-range values to $D_\mathrm{max}-1$, whereas MT-013 rejects them.
Both interfaces therefore expose the full \SI{4.069}{\nano\second} hardware delay grid.
Changing a delay selects a stored sample directly; fractional-delay interpolation is not implemented.

\subsubsection*{Memory cost and deployed configurations}
The sample-history storage in either variant is $2N_\mathrm{tap}D_\mathrm{max}r$ bits, with $r=16$ bits per component.
For the depths considered here, a 16-bit history maps to $k=\lceil D_\mathrm{max}/1024\rceil$ \gls{bram}18 equivalents, giving $2N_\mathrm{tap}k$ per filter, subject to synthesis mapping.
Both variants require 64 tap multipliers per channel. At depth 4096, the real and complex delay histories require 256 and 128 \gls{bram}18 equivalents, respectively; the complex variant trades half the tap count for complex arithmetic and smaller history storage. Its Doppler engine adds eight DSP slices and one BRAM36 per filter, excluding surrounding control logic.
Table~\ref{tab:resources} summarizes the resulting filter configurations; Section~\ref{sec:resources} addresses the remaining platform overhead and implementation limits.

\begin{table}[ht]
  \centering
  \caption{Deployed sparse FIR configurations and nominal per-channel resources.}
  \label{tab:resources}
  \small
  \begin{tabular}{lrr}
    \toprule
    Parameter & Real & Complex \\
    \midrule
    Tap slots $N_\mathrm{tap}$ & 32 & 16 \\
    Buffer depth $D_\mathrm{max}$ & 4096 & 4096 \\
    Maximum delay index & 4095 & 4095 \\
    Delay span ($\mu$s) & 16.663 & 16.663 \\
    Tap multipliers (DSP48) & 64 & 64 \\
    Delay storage (BRAM18 equiv.) & 256 & 128 \\
    Doppler engine DSP48 & 0 & 8 \\
    Doppler ROM (BRAM18 equiv.) & 0 & 2 \\
    Datapath stages $L$ & 9 & 9 \\
    \bottomrule
  \end{tabular}
    \vspace{-.3cm}
\end{table}

\subsection{Attenuation}
\label{sec:atten}

\lchem combines per-tap coefficient scaling with a programmable arithmetic right shift after each sparse \gls{fir}, following the two-stage approach of OpenAirLink~\cite{deshpande2024openairlink}.
The coefficients set the relative path gains, while the shift applies coarse attenuation to the entire channel.
This separation extends the nominal attenuation range and permits finer coefficient-based gain settings at large attenuation values.

\subsubsection*{Coefficient range and resolution}
Each coefficient component is a signed Q1.15 word, with $r=16$ total bits and $F=r-1=15$ fractional bits.
For a positive real coefficient code $b\in\{1,\ldots,2^F-1\}$ and a nonnegative shift $j$, the ideal combined gain is
\begin{equation}
  G(b,j)=20\log_{10}\!\left(\frac{b}{2^F}\right)
          -j\,20\log_{10}(2)\quad[\mathrm{dB}].
  \label{eq:gain}
\end{equation}
This expression neglects sample rounding, truncation, and saturation; the corresponding complex-path gain depends on the coefficient magnitude.
The positive coefficient codes alone span $20\log_{10}(2^F-1)\approx\SI{90.31}{\decibel}$, but their adjacent gain steps are nonuniform:
\begin{equation}
  \Delta G(b)=20\log_{10}\!\left(1+\frac{1}{b}\right).
  \label{eq:gain_res}
\end{equation}
The step is approximately \SI{6.02}{\decibel} at $b=1$ and decreases as the coefficient increases.
Small coefficient codes therefore provide coarse gain control even though they represent large attenuation.

\subsubsection*{Coarse attenuation with fine gain settings}
An arithmetic right shift by $j$ bits approximately scales each output I/Q component by $2^{-j}$, adding \SI{6.02}{\decibel} of attenuation per bit.
For a chosen maximum shift $s$, the ideal nonzero gain settings span
\begin{equation}
  D_\mathrm{nom}=20\log_{10}(2^F-1)+s\,20\log_{10}(2).
  \label{eq:drange}
\end{equation}
Within that range, assigning coarse attenuation to the shift allows larger coefficients to represent the remaining gain.
For a single positive tap kept in the interval $2^{F-1}\leq b\leq2^F-1$, adjacent coefficient steps satisfy
\begin{equation}
  \Delta G(b)\leq20\log_{10}\!\left(1+2^{-(F-1)}\right)
  \approx\SI{0.00053}{\decibel}.
  \label{eq:gbs}
\end{equation}
For example, $(b,j)=(32,0)$ and $(16384,9)$ both give an ideal attenuation of \SI{60.21}{\decibel}, but their adjacent coefficient steps are approximately \SI{0.267}{\decibel} and \SI{0.00053}{\decibel}, respectively.
The shift thus extends the range over which fine nominal gain settings are available.
This is useful when sweeping path loss near a receiver-sensitivity or link-adaptation threshold: assigning coarse attenuation to the shift leaves larger coefficients available for fine gain adjustments without changing the analog operating point.
These finer settings do not imply equivalent RF measurement accuracy.
In a multipath channel, the shift is shared by all taps; common coefficient scaling preserves their relative gains subject to quantization, while weaker taps can still have coarser relative resolution.

The nominal attenuation range is therefore $D_\mathrm{nom}\approx 90.31+6.02s$\,dB, with fine coefficient-based gain settings maintained by assigning coarse attenuation to the shift stage.
These values describe the digital gain settings; usable RF dynamic range is evaluated separately in Section~\ref{sec:verification}.
The implementation discards low-order bits during shifting and applies the shift after FIR saturation, so coefficient scaling must preserve sufficient headroom to avoid clipping before attenuation.

\subsection{Doppler Emulation}
\label{sec:doppler}

Doppler is produced by continuously rotating a path's complex coefficient. The rotation rate determines the frequency shift: a positive rate shifts the signal upward in frequency, and a negative rate shifts it downward. Each complex FIR contains one FPGA engine that generates these rotations for all its taps. The host supplies a \emph{channel snapshot}, specifying the path gains, delays, and Doppler frequencies; the engine then evolves the coefficients until the next snapshot arrives.

\emph{Per-path coefficient generation.}
Between snapshots, the desired coefficient of tap $i$ is
\begin{equation}
 b_i[n]=\mathcal{Q}_{\mathrm{Q1.15}}\!\left(
 \sum_{c=0}^{K-1} A_{i,c}e^{j2\pi f_{D,i,c}n/f_s}\right),
 \label{eq:doppler_rotation}
\end{equation}
where $A_{i,c}$ sets the magnitude and initial phase of component $c$, and $f_{D,i,c}$ is its Doppler frequency. One component represents a path with constant magnitude and a single Doppler shift. Summing several components at the same tap allows their relative phases to produce a time-varying magnitude, representing fading at that delay. The deployed image supports up to $K=4$ components per tap.

The FPGA stores phase in 32-bit accumulators. Each signed frequency word $\mathrm{FW}_{i,c}$ sets the phase increment per input sample, giving
\begin{equation}
 f_{D,i,c}=\mathrm{FW}_{i,c}\frac{f_s}{2^{32}},\qquad
 \Delta f_D=\SI{0.05722}{\hertz}.
\end{equation}
Thus, phase follows the number of samples accepted by the FIR rather than the timing of host writes. A shared arithmetic engine cycles through the 16 taps and their four components. It calculates each component once every 64 samples, advancing its phase by 64 increments on each visit to preserve the requested rotation rate.

\emph{Interpolation between coefficient updates.}
At $f_s=245.76$\,MS/s, the 64-sample cycle gives a coefficient refresh rate of $f_r=\SI{3.84}{\mega\hertz}$. Between calculated values, the FPGA linearly interpolates each coefficient, supplying a value to the FIR on every accepted sample. This method is called \emph{first-order hold} (FOH). It avoids the abrupt steps of \emph{zero-order hold} (ZOH), which keeps a coefficient constant until the next refresh. Unused components do not shorten the schedule.

Interpolation approximates the desired rotation and can produce unwanted spectral copies, called \emph{images}. For a single component, these copies occur at offsets of $\pm f_r,\pm2f_r,\ldots$ from the desired shifted tone. Their levels increase with Doppler magnitude. FOH also reduces the desired tone's amplitude by approximately $\operatorname{sinc}^2(f_D/f_r)$, where $\operatorname{sinc}(x)=\sin(\pi x)/(\pi x)$. Consequently, the refresh-rate Nyquist limit of $|f_D|<\SI{1.92}{\mega\hertz}$ is not a guarantee of low distortion throughout that range. Section~\ref{sec:verification_doppler} evaluates frequency error and unwanted image levels separately.

\emph{Applying channel snapshots.}
Three parameter banks allow the host to prepare a snapshot without modifying the channel currently in use. Host writes enter the shadow bank; a commit copies them to the staged bank; the engine then activates the complete snapshot at a 64-sample boundary. All components and pending delay changes within a filter activate together. A requested timestamp selects the first boundary at or after that time, with a granularity of \SI{260.4}{\nano\second}. A frequency change can retain the current phase or explicitly load a new phase. The reported application time marks parameter activation; the interpolated coefficients begin their transition one frame later and reach the new endpoint after a second frame. The FIR delay buffers, sample multipliers, and RFNoC routing are unchanged.

\subsection{Runtime Control}
\label{sec:runtime_control}

External scenario software streams binary control messages over ZeroMQ to the emulator application on the X410's embedded CPU (Fig.~\ref{fig:arch}).
Channel-profile messages update the per-link tapped-delay \gls{cir} within the loaded image's tap-count and delay limits, while separate messages control per-link coarse attenuation.
The application maps requests to the six directional channels and writes their FPGA registers through UHD/RFNoC.
Delays are carried as native sample indices, exposing the full hardware delay grid described in Section~\ref{sec:delay}.
External propagation models can therefore update channel conditions without rebuilding the FPGA image.

Radio configuration is controlled separately through MT-001, which sets receive and transmit center frequencies and analog gains for the four RF interfaces through UHD.
A shared-memory interface on the X410 also accepts these settings from its dashboard service and publishes values read back from the radios.
These settings control carrier frequency and RF levels, while FIR coefficients and right shifts determine the emulated channel response.

MT-013 carries native tap delays, complex component gains, frequency words, starting phases, phase-retention flags, and an optional device timestamp. The application prepares each addressed filter before issuing its commit. A common future timestamp coordinates links subject to their frame boundaries and timely delivery; untimed commits are local to each filter. Sionna~RT can supply one ray per tap through MT-013 while retaining phase across snapshots. MT-012 continues to accept directly specified coefficients. RX timestamps provide the commit time base; FIR output timestamps are stripped so that downstream transmitters do not reject processed packets as late. Snapshot delivery determines how rapidly scenario parameters can change, while the FPGA autonomously generates the intervening Doppler evolution.

\section{Resource Management}
\label{sec:resources}

\subsection{Complex-Tap Resource Budget}

The USRP X4x0 series (X410, X420, and X440) shares the ZU28DR fabric, with 4{,}272 DSP48E2 slices and 2{,}160 \gls{bram}18 equivalents~\cite{x4x0HardwareManual,amd_ds890}. A 16-tap complex filter with 4{,}096-sample histories uses nominally 64 DSP slices and 128 \gls{bram}18 equivalents (Table~\ref{tab:resources}). Including eight DSP slices and two \gls{bram}18 equivalents for each Doppler engine, for $N_\mathrm{ch}$ directional filters,
\begin{equation}
  \begin{aligned}
  D_\mathrm{total} &= D_\mathrm{plat} + 72N_\mathrm{ch},\\
  B_\mathrm{total} &= B_\mathrm{plat} + 130N_\mathrm{ch},
  \end{aligned}
  \label{eq:dsp_bram_total}
\end{equation}
where the platform terms depend on the target image. Using the implementation's analytical baseline of $D_\mathrm{plat}=2{,}791$ and $B_\mathrm{plat}=210$, the six-filter configuration totals 3{,}223 DSP slices (75.4\%) and 990 \gls{bram}18 equivalents (45.8\%). Reference budgets of 80\% DSP and 90\% \gls{bram} permit eight such filters under these assumptions; this is an analytical estimate, not a timing guarantee. The routed Doppler image actually uses 3{,}335 DSP slices (78.07\%), 604 BRAM36 tiles (55.93\%), 259{,}235 LUTs (60.96\%), and all 80 URAMs. These totals include the complete platform and control logic omitted by the simplified model.

The 16-tap complex configuration balances the number of resolvable paths against delay span and parallel link count. Doubling the tap count to 32 at the same delay depth doubles the FIR's multiplier and history-memory costs. The 16-tap default retains a \SI{16.66}{\micro\second} delay span while leaving more resources available for additional channel filters and the surrounding image.

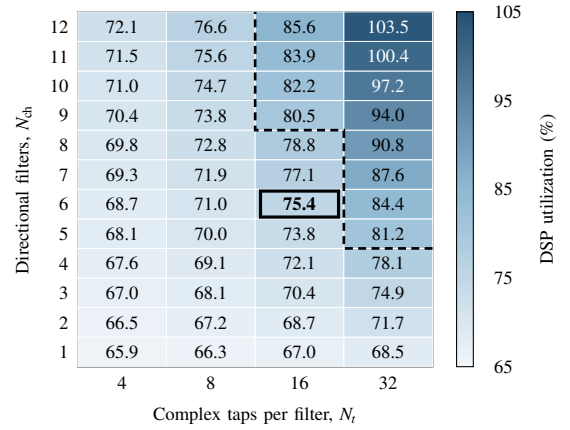
\begin{figure}[t]
    \centering
    \input{Figures/fpga_dsp_heatmap}
    \caption{Modeled DSP utilization (\%) for complex FIR filters. The lower-left region meets the 80\% reference budget; the outlined cell marks the deployed configuration.}
    \label{fig:dsp_heatmap}
    \vspace{-.3cm}
\end{figure}

Fig.~\ref{fig:dsp_heatmap} illustrates the DSP tradeoff between tap count and filter count. More taps increase both DSP and memory demand, whereas deeper histories increase memory without adding multipliers. The modeled DSP cost per filter is $4N_t+8$, and its memory cost is $2N_t\lceil D_\mathrm{max}/1024\rceil+2$ \gls{bram}18 equivalents; sufficiently long delays can therefore make memory the limiting resource.

These costs distinguish two scaling choices. Reducing the synthesized tap count releases both DSP and memory resources, allowing more independently filtered links when fewer paths per link are sufficient. Reducing delay-buffer depth releases only memory, so shortening the delay span alone does not increase the eight-filter estimate while DSP remains limiting. Both changes require a new FPGA image; disabling taps at runtime does not reclaim their allocated hardware.

\subsection{Fabric Scaling and RF Connectivity}

The tap-count tradeoff in Fig.~\ref{fig:dsp_heatmap} determines which link topologies fit within the fabric budget: $N_\mathrm{ch}$ counts independently filtered directional links, not RF ports. For the default 16-tap complex filter, Eq.~\eqref{eq:dsp_bram_total} gives
\begin{equation}
 N_{\mathrm{ch},\max}=\min\!\left(
 \left\lfloor\frac{D_\mathrm{budget}-D_\mathrm{plat}}{72}\right\rfloor,
 \left\lfloor\frac{B_\mathrm{budget}-B_\mathrm{plat}}{130}\right\rfloor
 \right).
\end{equation}
A bidirectional star with $P$ RF nodes requires $2(P-1)$ filters, while a fully connected directed topology requires $P(P-1)$.
For four nodes, the star in Fig.~\ref{fig:arch} uses six filters and 75.4\% of the DSP slices under the preceding baseline assumptions. A full mesh requires twelve filters, raising modeled DSP utilization to 85.6\%, above the 80\% reference budget (Fig.~\ref{fig:dsp_heatmap}). Reducing each filter to eight complex taps brings the twelve-link estimate back to 76.6\%, trading paths per link for connectivity.
The X410, X420, and X440 expose four, two, and eight full-duplex RF channels, respectively~\cite{x4x0HardwareManual}; RF-port availability therefore further restricts the realizable topology. These estimates retain the same platform baseline; larger graphs require adapted routing and control mappings, with resource use and timing verified for the target image and sample rate.

\section{Verification}
\label{sec:verification}

We validate the \lchem implementation on a \gls{usrp} X410 by applying controlled test waveforms and measuring the resulting RF signals. The experiments characterize delay, attenuation, multipath fidelity, multi-channel operation, Doppler emulation, and four-component fading. Table~\ref{tab:verification_plan} summarizes the tests and identifies the design property evaluated by each. Each experiment is repeated 10 times, and reported results are averaged across repetitions unless otherwise stated.

\begin{table}[t]
  \centering
  \caption{Verification plan.}
  \label{tab:verification_plan}
  \scriptsize
  \setlength{\tabcolsep}{3pt}
  \begin{tabularx}{\columnwidth}{@{}ll>{\raggedright\arraybackslash}Xl@{}}
    \toprule
    \textbf{ID} & \textbf{Claim} & \textbf{Test} & \textbf{Subsec.}\\
    \midrule
    E1 & Delay accuracy and span        & Tap-index sweep with matched-filter delay & V-B\\
    E2 & Attenuation accuracy           & Joint coefficient + shiftright sweep         & V-B\\
    E3 & Multipath CIR fidelity         & Wiener deconvolution on programmed PDPs      & V-B\\
    E4 & 4-ch topology linearity        & DL match and UL triple-input combine         & V-C\\
    E5 & Static complex coefficients    & Phase $\{45,90,180,270\}^\circ$, $f_d=0$     & V-D\\
    E6 & Doppler accuracy and purity    & CW phase slope and spectral images   & V-D\\
    E7 & Snapshot timing                & Phase retention and timestamped commits                   & V-D\\
    E8 & Multipath and fading           & Delay--Doppler recovery and four-component ACF & V-D\\
    \bottomrule
  \end{tabularx}
\vspace{-.3cm}
\end{table}

\subsection{Setup}
\label{sec:verification_setup}

\begin{figure}
    \centering
    \input{Figures/lchem_validation}
    \vspace{-.5cm}
    \caption{Verification setup. 
    }
    \label{fig:verification}
    \vspace{-.3cm}
\end{figure}
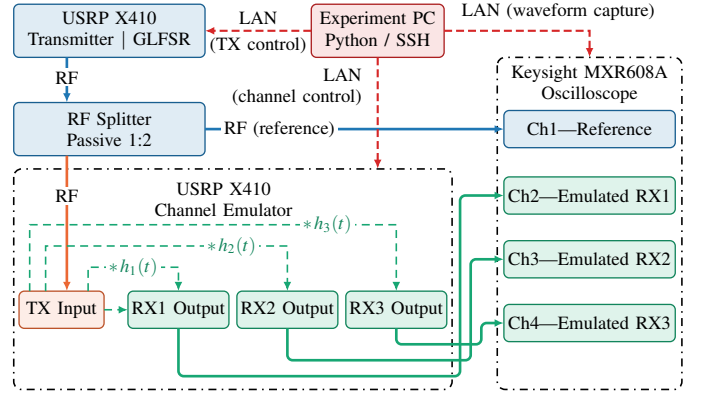

The verification setup is illustrated in Fig.~\ref{fig:verification}. An external experiment PC coordinates device control and oscilloscope acquisition; the emulator application runs on the X410's embedded CPU. A second X410 generates a degree-16 \gls{glfsr} \gls{bpsk} sounder at \SI{3.55}{\giga\hertz}, replayed continuously.
The $2^{16}-1=65{,}535$-sample period spans \SI{266.7}{\micro\second} at the native \SI{245.76}{\mega Sa\per\second} IQ rate, more than an order of magnitude longer than either \gls{fpga} variant's maximum delay span, so the recovered \gls{cir} suffers no period-aliasing within the emulator's operating range.

A passive splitter routes the sounder signal to both channel~1 of a Keysight MXR608A real-time oscilloscope and the emulator’s gNB input (radio0:RX0). The oscilloscope records the three downlink outputs (radio0:TX1, radio1:TX0, and radio1:TX1) on channels~2--4 over a \SI{10}{\milli\second} window.
X410s lock to a shared NI/Ettus OctoClock CDA-2990 distributing \SI{10}{\mega\hertz} and PPS references, and all devices are sharing same reference signal. Static-channel test points are captured ten times for variance analysis; the Doppler campaign uses the setup and repetition counts specified in Section~\ref{sec:verification_doppler}.


\textit{Calibration.} Five pass-through runs with identity \glspl{fir}, with the emulator re-initialised between runs, yield a mean end-to-end SMA-to-SMA latency of \SI{2.483}{\micro\second} on DL0 ($\sigma = \SI{1.5}{\nano\second}$). Cable skew is removed using a barrel through-connection. DL1 and DL2 lag DL0 by \SI{16.7}{\nano\second} and \SI{16.2}{\nano\second}, respectively. The RF level offsets relative to the reference are \SI{12.9}{\decibel}, \SI{14.3}{\decibel}, and \SI{8.2}{\decibel} for DL0, DL1, and DL2, respectively ($\sigma \leq \SI{0.12}{\decibel}$). These baselines are subtracted from subsequent measurements so that the reported errors reflect emulator configuration accuracy after correction for fixed hardware offsets.

\textit{Post-processing.} Delay is recovered as the lag of the peak of the normalized cross-correlation between the Ch1 reference and each emulated output.
Attenuation is the RMS-power ratio (in dB) between the emulated output and Ch1, with the calibration baseline removed. Multipath \glspl{cir} are obtained by Wiener deconvolution of the DL1 output against the DL0 pass-through reference,
\begin{equation}
    \hat{H}(f) = \frac{Y(f)\,X^{*}(f)}{|X(f)|^{2} + \lambda},\qquad
    \hat{h}(t)=\mathcal{F}^{-1}\{\hat{H}(f)\},
    \label{eq:wiener}
\end{equation}
with $\lambda = 10^{-4}\,\max|X(f)|^{2}$ a Tikhonov term that suppresses noise amplification at probe-spectrum nulls. 

\subsection{Static Channel}
\label{sec:verification_static}

\subsubsection*{E1 -- Delay linearity}
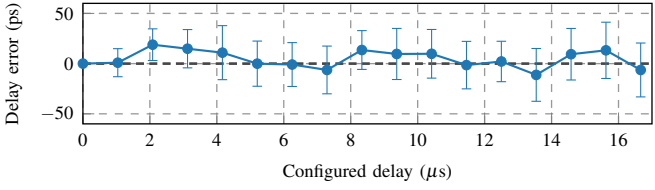
\begin{figure}[t]
    \centering
    \begingroup
    \def\lchemPanelWidth{0.85\columnwidth}
    \input{Figures/delay_error}
    \endgroup
    \caption{E1: delay error over tap indices 0--4095 (\SIrange{0}{16666.624}{\nano\second}) after
    per-sweep latency calibration. Markers show the mean over 10 full sweeps; error bars show $\pm1$ standard deviation across sweeps. The dashed line marks zero error.}
    \label{fig:delay_verification}
    \vspace{-.3cm}
\end{figure}

The configured tap index was swept over $17$ settings from $0$ to $4095$ (in steps of $256$, plus the last tap), spanning the full delay line from \SI{0}{\nano\second} to \SI{16.66}{\micro\second} at $\Delta\tau_\mathrm{tap}=\SI{4.069}{\nano\second}$, on the complex-16 image. The sweep was repeated $10$ times (170 captures, all valid). Each delay is estimated as the envelope peak, with parabolic sub-sample interpolation, of the matched-filter \gls{cir} against the Ch1 reference, after subtracting the pass-through latency (\SI{2486.87}{\nano\second}, \SI{0.02}{\nano\second} standard deviation) calibrated from the tap-0 capture of the same sweep. The averaged result recovers the programmed delay with \textbf{RMS \SI{0.0095}{\nano\second}} and peak \SI{0.019}{\nano\second} error (Fig.~\ref{fig:delay_verification}); the fitted slope is \SI{4.069009}{\nano\second\per tap} against the nominal \SI{4.069010}{\nano\second\per tap}, with an intercept of \SI{0.008}{\nano\second}. The sweep-to-sweep standard deviation is \SIrange{0.014}{0.028}{\nano\second} per setting, so the residual is at the measurement noise floor, far below the native tap quantum, and the delay is linear over the entire \SI{16.66}{\micro\second} line.

\subsubsection*{E2 -- Attenuation accuracy}

The emulator controls attenuation through two stages (Section~\ref{sec:atten}): \gls{fir} coefficient scaling and \texttt{shiftright} block. E2 measures the attenuation accuracy of each stage separately and in combination.

\begin{figure}[t]
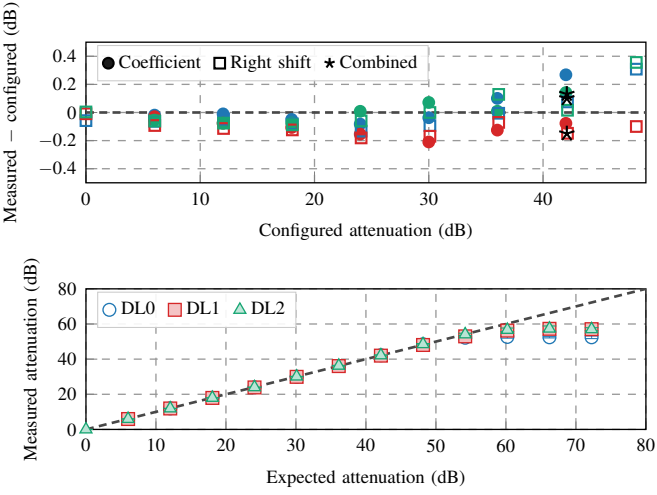

    \centering
    \lchemStack{Figures/attenuation_error,Figures/attenuation_bitshift}
    \caption{E2: attenuation accuracy. Upper plot: signed error for
    coefficient scaling, right shifting, and their combination across
    three downlinks; the combined setting targets approximately
    \SI{42}{\decibel}. Lower plot: measured right-shift attenuation
    over the full \SIrange{0}{72}{\decibel} sweep. The response departs
    from the ideal line above approximately \SI{50}{\decibel} as the
    output approaches the measurement noise floor.}
    \label{fig:att_verification}
    \vspace{-.3cm}
\end{figure}

\textit{Stage~A (coefficient scaling).} We sweep the leading
\gls{fir} coefficient from $2^{14}$ to $2^{8}$, corresponding to
approximately \SIrange{6}{42}{\decibel} of programmed attenuation
in \SI{6}{\decibel} steps. After subtracting the calibration
baseline, the error across the three downlinks has an RMS value
of \SI{0.11}{\decibel} and a maximum magnitude of
\SI{0.27}{\decibel} (Fig.~\ref{fig:att_verification}, upper plot).
The capture-to-capture standard deviation increases with
attenuation, reaching \SI{0.33}{\decibel} at the highest setting.

\textit{Stage~B (right shift).} With the coefficient at full
scale, we sweep the shift amount from $s=0$ to $s=12$, giving
a nominal attenuation of approximately \SIrange{0}{72}{\decibel}
in \SI{6.02}{\decibel} steps. The measured error remains within
\SI{0.36}{\decibel} through \SI{48}{\decibel}. At higher settings,
the measured attenuation falls increasingly below the programmed
value as the output approaches the oscilloscope noise floor
(Fig.~\ref{fig:att_verification}, lower plot).

\textit{Combined (A+B).} We apply approximately
\SI{18}{\decibel} through coefficient scaling and
\SI{24}{\decibel} through a four-bit right shift, targeting
\SI{42.08}{\decibel} in total. Across the three downlinks, the
measured attenuation differs from this target by at most
\SI{0.15}{\decibel}. The stars in the upper plot of
Fig.~\ref{fig:att_verification} show this combined setting.

\subsubsection*{E3 -- Multipath CIR fidelity}
\begin{figure}[!t]
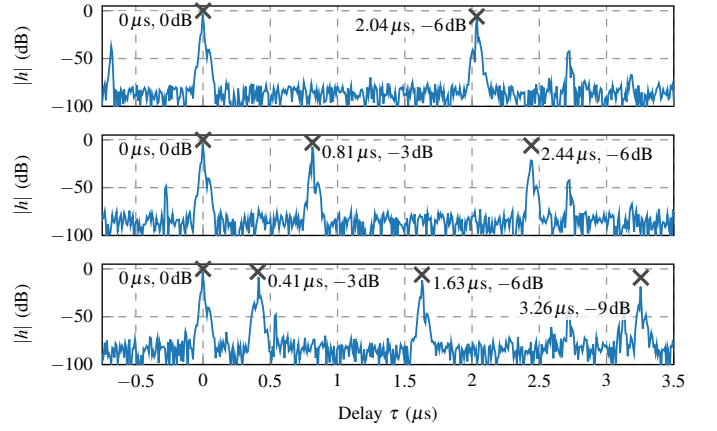

    \centering
    \lchemStack{Figures/cir_mp_2tap,Figures/cir_mp_3tap,Figures/cir_mp_4tap}
    \caption{E3: measured CIRs for three programmed PDPs using a sparse
         \gls{fir} with tap spacing
         $\Delta\tau_\mathrm{tap}=\SI{4.069}{\nano\second}$.
         Blue lines show the measured CIRs, and cross markers indicate
         the detected peaks, annotated with timing and gain deviations
         $(\Delta t,\Delta G)$.}
    \label{fig:cir_sparse}
    \vspace{-.3cm}
\end{figure}

Three power-delay profiles are programmed and recovered via Wiener deconvolution in Eq.~\eqref{eq:wiener}: a two-tap profile at index~500 ($\Delta\tau\!\approx\!\SI{2035}{\nano\second}$, secondary at $-6$\,dB); a three-tap exponential decay at indices $\{0,200,600\}$ with gains $\{0,-6,-12\}$\,dB; and a four-tap profile at indices $\{0,100,400,800\}$ spanning \SI{3.26}{\micro\second}. Detected peaks are flagged where $|\hat{h}(t)|$ exceeds the noise floor by $>$\SI{10}{\decibel} with a minimum prominence of \SI{8}{\decibel}. All programmed taps are resolved at their target positions (Fig.~\ref{fig:cir_sparse}). 

\subsection{Multi-channel Topology}
\label{sec:verification_topology}

Experiments E1--E3 characterize a single \gls{fir} path. E4 verifies that
the 4-channel split--combine topology of
Section~\ref{sec:rfnoc_signal_graph} preserves linearity end-to-end.

\subsubsection*{E4a -- Downlink match} 
With identity \gls{fir}s on all three downlinks, the gain and delay of DL0/DL1/DL2 relative to the splitter input are measured. 
The result is the per-channel calibration baseline used throughout this section
(\SIrange{5.6}{6.1}{\decibel}; \SI{62.5}{\nano\second} and
\SI{45.0}{\nano\second} pipeline offsets relative to DL0). These
offsets are stable to $<$\SI{0.1}{\decibel} and
$<$\SI{1}{\nano\second} across acquisitions, demonstrating that the
cascaded splitter introduces only a fixed pipeline imbalance, not a
variable one.

\subsubsection*{E4b -- Uplink combining}
\begin{figure}[t]
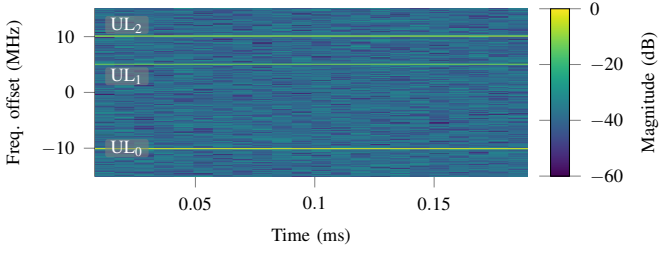

    \centering
    \lchemStack{Figures/ul_combine_linearity}
    \caption{E4b: spectrogram of the combined uplink output (scope Ch1)
         with all three UL \gls{fir}s set to unity gain, showing
         steady tones at $-10$, $5$, and $10$\,\si{\mega\hertz}.}
    \label{fig:ul_combine_linearity}
    \vspace{-.3cm}
\end{figure}

The uplink path sums the outputs of three \gls{fir}s through an addsub cascade and routes the result to radio0:TX0. 
We test the combiner using transmit channels 0, 1, and 2 of a companion X410, each connected by a separate cable to an UL input. 
A single \texttt{tx\_streamer} drives all three channels to keep the branches sample-aligned. 
Each channel transmits a continuous-wave tone at a distinct baseband offset:
$(f_0,f_1,f_2)=(-10,5,10)\,\si{\mega\hertz}$. With all three UL \gls{fir}s configured for unity gain and no right shift applied, the spectrogram of the combined output (Fig.~\ref{fig:ul_combine_linearity}) shows three steady horizontal lines at the expected frequencies.

\subsection{Complex Coefficients and Doppler}
\label{sec:verification_doppler}

Experiments E5--E8 evaluate the FPGA Doppler image with four components per tap. This campaign uses a B200mini downlink source and a B210 uplink source at \SI{3.5375}{\giga\hertz}; the oscilloscope's \SI{10}{\mega\hertz} reference is distributed to the radios through the OctoClock. CW probes have a \SI{2}{\mega\hertz} baseband offset. The oscilloscope measures the source reference and DL0, with DL1 or the combined uplink output included as required. Tone sweeps average three captures per setting; the spectral comparison uses one capture per mode.

\subsubsection*{E5 -- Static complex coefficients}
We first verify that the complex coefficients preserve signal magnitude when their phase is fixed. With the Doppler engine disabled, four CW measurement cycles at $45$, $90$, $180$, and $270$ degrees give level differences within $\pm\SI{0.01}{\decibel}$ of neighboring pass-through captures. The 95\% confidence-interval half-widths are at most \SI{0.04}{\decibel}. These measurements establish magnitude preservation. Absolute RF phase is affected by reference-to-output offsets in the measurement chain and requires separate calibration.

\subsubsection*{E6 -- Doppler accuracy and spectral purity}
\emph{Frequency accuracy.}
We apply a CW probe and sweep a single component over $f_D\in\{\pm1,\pm10,\pm100,\pm10^3,\pm10^4,\pm10^5,\pm10^6\}$\,Hz and $+200$\,Hz. Each measurement uses a \SI{20}{\milli\second} capture. After digital downconversion, multiplying the output by the conjugate reference gives a beat signal whose phase rotates at the Doppler frequency. We use the FFT for an initial frequency estimate, then fit the slope of the beat signal's unwrapped phase.

The inverse capture duration is \SI{50}{\hertz}, but the phase-slope estimate is not restricted to FFT-bin centers. For example, \SI{1}{\hertz} and \SI{10}{\hertz} shifts produce $7.2^\circ$ and $72^\circ$ of phase rotation over the capture. These rotations support single-tone frequency estimation below \SI{50}{\hertz}; they do not establish the ability to resolve two closely spaced tones. After subtracting the carrier-offset baseline measured at zero Doppler, the largest absolute mean error across three captures per setting is \SI{0.0311}{\hertz}. This error is relative to the frequency represented by the programmed word; rounding the requested frequency to that word contributes at most another \SI{0.0286}{\hertz}.

\begin{figure}[!t]
 \centering
 \includegraphics[width=\columnwidth]{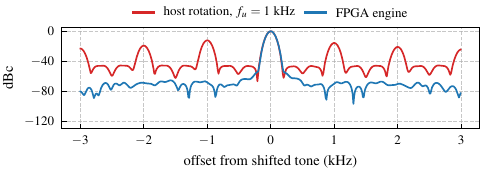}
 \caption{E6: measured spectra at a \SI{200}{\hertz} Doppler shift, normalized to the shifted tone. FPGA coefficient synthesis suppresses the staircase spurs of the \SI{1}{\kilo\hertz} host-update baseline.}
 \label{fig:dop:fpga_host}
 \vspace{-2mm}
\end{figure}

\emph{Suppression of host-update spurs.}
Figure~\ref{fig:dop:fpga_host} compares the FPGA engine with a \SI{1}{\kilo\hertz} host-update reference at $f_D=\SI{200}{\hertz}$. Holding coefficients between host writes creates a staircase rotation and unwanted tones, or spurs, around the desired signal. The strongest such spur is $-12.1$\,dBc, whereas the worst FPGA near-carrier spur is $-65.4$\,dBc: a measured improvement of \SI{53.3}{\decibel}, with a floor-limited lower bound of \SI{50.8}{\decibel}. Here, dBc expresses power relative to the desired shifted tone; a more negative value indicates a weaker unwanted component.

\begin{figure}[!t]
 \centering
 \includegraphics[width=\columnwidth]{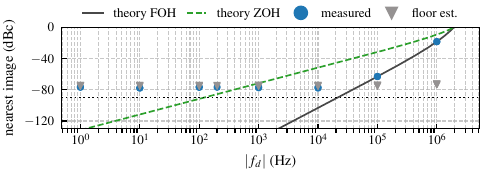}
 \caption{E6: unwanted spectral-copy levels versus Doppler magnitude. Linear interpolation (FOH) suppresses images more strongly than constant-value updates (ZOH). At small shifts, measurements are limited by the bench floor.}
 \label{fig:dop:images}
 \vspace{-2mm}
\end{figure}

\emph{Spectral limits at larger Doppler shifts.}
Figure~\ref{fig:dop:images} measures the interpolation images defined in Section~\ref{sec:doppler}. Its horizontal axis is Doppler magnitude, and its vertical axis is the power of the strongest first image relative to the desired tone. The FOH curve predicts the behavior of the deployed linear interpolation; the ZOH curve shows the larger image levels expected from holding coefficients constant between refreshes.

At low Doppler shifts, the predicted FOH images lie below the approximately $-74$\,dBc measurement floor. The measured points there indicate the bench's sensitivity rather than the actual image level. At $|f_D|=\SI{100}{\kilo\hertz}$, the first images rise to approximately $-63$\,dBc. At $|f_D|=\SI{1}{\mega\hertz}$, the strongest reaches $-18.1$\,dBc, and the desired tone loses about \SI{2}{\decibel}, consistent with the interpolation model. Thus, the engine retains accurate frequency placement at large shifts, but the resulting signal contains stronger unwanted copies. Other wideband spurs near $-51$\,dBc also appear in the source reference, preventing this bench from establishing a $-90$\,dBc RF spur floor.

\emph{Multiple links and uplink operation.} Simultaneous DL0/DL1 shifts spanning $-\SI{50}{\kilo\hertz}$ to $+\SI{100}{\kilo\hertz}$ have absolute mean errors at most \SI{0.030}{\hertz}, using a separate carrier-offset baseline for each link. The other link's tone appears between $-67.8$ and $-78.5$\,dBc, including measurement-floor-limited cases. Uplink measurements through the combiner at $\{-1000,-100,100,1000,100000\}$\,Hz give absolute mean errors at most \SI{0.0229}{\hertz}.

\subsubsection*{E7 -- Phase retention and timed snapshots}
This experiment checks whether a snapshot can change frequency without resetting phase, and whether a timed update activates within the specified frame. We alternate between \SI{100}{\hertz} and \SI{110}{\hertz} every \SI{50}{\milli\second}, retaining the accumulated phase. Segmented CW acquisitions span \SI{1.392}{\second}, covering approximately 27 changes. The largest observed phase-increment residual is $1.9^\circ$, below the $5^\circ$ detection threshold, with $0.6^\circ$ RMS measurement noise. The gaps between acquisitions could conceal a phase step of up to $12.2^\circ$, which limits the continuity claim. For comparison, explicitly reloading phase produces 19 detected steps, confirming that the measurement detects phase resets.

Three timestamped updates activate 7, 39, and 44 sample ticks after the requested time, corresponding to approximately 28, 159, and \SI{179}{\nano\second}. All fall within the specified 64-sample frame. These register readbacks verify parameter activation; the coefficient interpolation transition follows as described in Section~\ref{sec:doppler}.

\subsubsection*{E8 -- Multipath Doppler and four-component fading}
\emph{Independent Doppler shifts on delayed paths.} A four-tap channel uses delay indices $\{0,61,246,982\}$, amplitudes $\{0.35,0.25,0.18,0.10\}$, and Doppler shifts $\{0,150,-300,400\}$\,Hz. A \SI{30.72}{\mega Sa\per\second} QPSK m-sequence probe supplies one \SI{266.6}{\micro\second} burst per millisecond; per-burst channel estimates recover all four delay--Doppler components. Mean Doppler errors relative to tap zero are at most \SI{0.03}{\hertz}. Estimating each path delay from the energy-weighted center of its recovered peak gives relative delay errors within \SI{0.5}{\nano\second}; peak-based estimates are sensitive to the probe's autocorrelation sidelobes. Unwanted response outside the programmed paths, near $-21$\,dB is also present in pass-through measurements and is a bench limitation.

\begin{figure}[!t]
 \centering
 \includegraphics[width=\columnwidth]{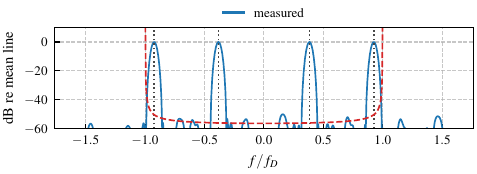}
 \caption{E8: four-component fading at $f_D=\SI{1000}{\hertz}$. The discrete commanded components sample the Jakes frequency distribution; the measured spectrum validates their synthesis.}
 \label{fig:dop:fading}
 \vspace{-2mm}
\end{figure}

\emph{Fading within one tap.} We next sum four rotating components at the same delay. Their changing relative phases cause constructive and destructive interference, making the tap magnitude vary over time. The components have equal amplitudes, with random initial phases and frequencies $f_c=f_D\cos[(2c+1)\pi/8]$, $c=0,\ldots,3$, for $f_D\in\{200,1000\}$\,Hz. The \SI{40}{\milli\second} captures recover all four spectral lines with maximum frequency errors of \SI{0.04}{\hertz} and \SI{0.02}{\hertz}, respectively, and line-power spreads of \SI{0.01}{\decibel} and \SI{0.02}{\decibel}. The normalized complex \gls{acf}, which measures how the channel remains correlated over time, differs from the programmed model by at most 0.034 and 0.006 for $\tau\leq1/f_D$. Figure~\ref{fig:dop:fading} shows the four measured lines at their programmed frequencies. Their locations sample the Jakes Doppler distribution, and their sum approximates its \gls{acf} over short lags. Four discrete components do not reproduce a continuous Jakes spectrum; these measurements therefore validate the programmed component sum rather than Rayleigh or Rician envelope statistics.

\section{System-Level Evaluation} 
\label{sec:ran}

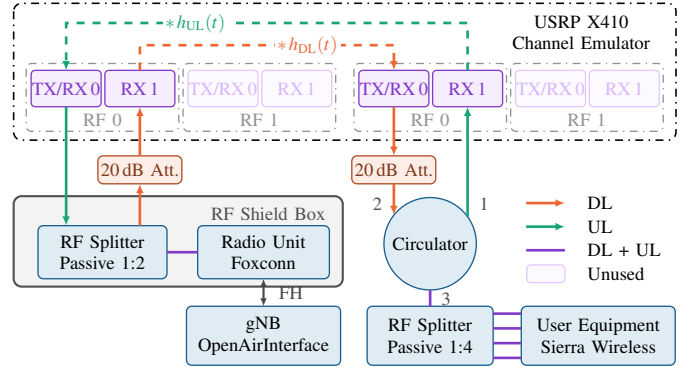
\begin{figure}[ht]
    \centering
    \input{Figures/experimental_setup}
    \vspace{-.5cm}
    \caption{5G RAN Evaluation Setup.}
    \label{fig:ran_topology}
    \vspace{-.3cm}
\end{figure}

Section~\ref{sec:verification} establishes channel-level accuracy. We now evaluate \lchem as the propagation medium of an operational \gls{ran}, whose radios adapt modulation, coding, and transmit power to the emulated channel. A commercial Foxconn \gls{ru} is driven by the \gls{oai} \gls{cu}/\gls{du} stack, with a Sierra Wireless EM9293 5G modem as the \gls{ue}. The modem sits in an RF shielding enclosure and connects by cable to prevent over-the-air leakage.

\subsection{Gain Calibration and Operating-Point Selection}
\label{sec:ran_dynrange}

The operating point depends on the emulator's dynamic range and the \gls{ran}'s power control and link adaptation. The \gls{ru} settings, fixed cable attenuators, and emulator analog gains jointly determine signal levels (Fig.~\ref{fig:ran_topology}). We therefore sweep output gain before selecting the configuration used in subsequent experiments.

\begin{figure}[ht!]
    \centering
    \begingroup
    \def\lchemPanelWidth{0.70\columnwidth}
  \sbox\lchemBoxA{\tikzset{every picture/.append style={lchem measure}}\input{Figures/ran_gain_snr_rsrp}}%
  \setlength\lchemAL{\lchemOverL}\setlength\lchemAW{\lchemAxisW}\setlength\lchemAR{\lchemOverR}
  \sbox\lchemBoxB{\tikzset{every picture/.append style={lchem measure}}\input{Figures/ran_gain_ul_uetx_pgf}}%
  \setlength\lchemBL{\lchemOverL}\setlength\lchemBW{\lchemAxisW}\setlength\lchemBR{\lchemOverR}

    \setlength{\lchemCA}{\lchemAL}
    \lchemMaxTo\lchemCA{\lchemBL}
    \setlength{\lchemCB}{\lchemAR}
    \lchemMaxTo\lchemCB{\lchemBR}
    \edef\lchemPanelWidth{%
      \the\dimexpr\linewidth-\lchemCA-\lchemCB-1pt\relax}
  \sbox\lchemBoxA{\tikzset{every picture/.append style={lchem measure}}\input{Figures/ran_gain_snr_rsrp}}%
  \setlength\lchemAL{\lchemOverL}\setlength\lchemAW{\lchemAxisW}\setlength\lchemAR{\lchemOverR}
  \sbox\lchemBoxB{\tikzset{every picture/.append style={lchem measure}}\input{Figures/ran_gain_ul_uetx_pgf}}%
  \setlength\lchemBL{\lchemOverL}\setlength\lchemBW{\lchemAxisW}\setlength\lchemBR{\lchemOverR}

  \par\noindent\makebox[\linewidth]{\input{Figures/ran_gain_legend}}\par
  \global\lchemAfterLegendtrue
    \global\lchemAfterLegendfalse
    \vspace{\lchemLegendGap}
    \vspace{-.2cm}

    \begin{subfigure}{\columnwidth}
        \noindent\makebox[\linewidth][l]{%
            \hspace*{\dimexpr\lchemCA-\lchemAL\relax}%
            \usebox{\lchemBoxA}\hss}
        \caption{Downlink \gls{sinr} (left axis)
        and \gls{rsrp} (right axis).}
        \label{fig:ran_gain_dl}
    \end{subfigure}

    \par\medskip
    \vspace{-.4cm}

    \begin{subfigure}{\columnwidth}
        \noindent\makebox[\linewidth][l]{%
            \hspace*{\dimexpr\lchemCA-\lchemBL\relax}%
            \usebox{\lchemBoxB}\hss}
        \caption{\gls{ue} transmit power.}
        \label{fig:ran_gain_uetx}
    \end{subfigure}
    
    \caption{Effect of emulator output-gain offsets across 
    $\pm\SI{10}{\decibel}$: (a) downlink \gls{sinr} 
    and \gls{rsrp}; (b) \gls{ue} transmit power.}
    \label{fig:ran_gain}
    \endgroup
\end{figure}
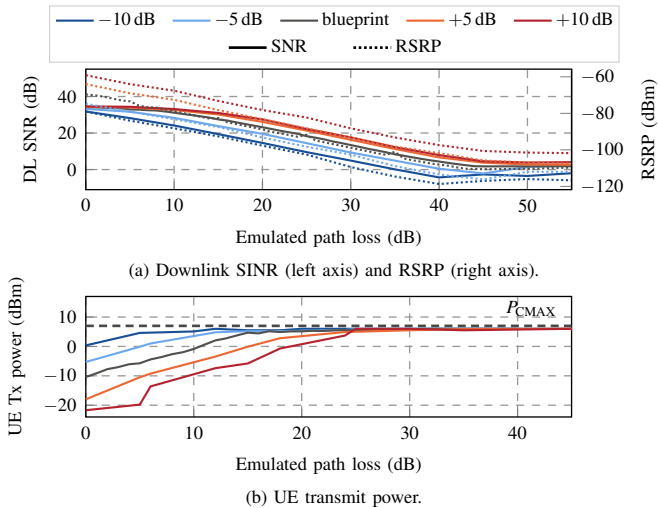

Fig.~\ref{fig:ran_gain} shows the response to emulator output-gain offsets from $-\SI{10}{\decibel}$ to $+\SI{10}{\decibel}$ relative to the baseline setting, in \SI{5}{\decibel} steps. 
For each downlink gain setting, we adjust \texttt{ssPBCH\_BlockPower} accordingly to keep the \gls{ue}'s path-loss estimate consistent with the configured downlink reference-signal power.
We evaluate each gain setting across the tested range of added path loss.

\emph{Downlink response.}
The measured \gls{rsrp} follows the output-gain increments at each added path loss (Fig.~\ref{fig:ran_gain_dl}, dotted curves).
The corresponding improvement in \gls{sinr}, however, depends on the received signal level.
At \SI{0}{\decibel} added path loss, increasing output gain by \SI{20}{\decibel} raises the \gls{sinr} from \SI{31.8}{\decibel} to \SI{34.6}{\decibel}, an increase of only \SI{2.8}{\decibel}.
The three highest gain settings, spanning \SI{10}{\decibel}, produce \gls{sinr} values within \SI{0.4}{\decibel} of one another. 
At \SI{20}{\decibel} added path loss, the \gls{sinr} spread across the five settings increases to \SI{12.9}{\decibel}. 
Thus, additional output gain provides little improvement near the observed \SI{34.6}{\decibel} ceiling but has a larger effect as path loss
increases.

This behavior is consistent with noise or distortion
present before the emulator's output-gain stage.
\lchem captures the received waveform together
with noise within its processing bandwidth.
Increasing output gain scales both the signal
and the noise already present at that stage.
Once these contributions dominate over the
\gls{ue}'s receiver noise, further output gain
provides little \gls{sinr} improvement.
This mechanism is common to RF channel emulators
that capture and replay signals.
However, the gain sweep alone does not isolate
the cause of the measured ceiling, which may
also reflect impairments in the \gls{ru},
emulator, or \gls{ue}.

\emph{Uplink response.}
Uplink power control compensates for changes in emulator output gain by adjusting the \gls{ue}'s transmit power (Fig.~\ref{fig:ran_gain_uetx}).
At \SI{0}{\decibel} added path loss, the \gls{ue} transmits at $-21.7$\,dBm with a $+\SI{10}{\decibel}$ gain offset and at $+0.4$\,dBm with a $-\SI{10}{\decibel}$ offset.
The \SI{20}{\decibel} gain increase therefore reduces the measured transmit power by \SI{22.1}{\decibel}.
As added path loss increases, the \gls{ue} raises its transmit power until it reaches $P_\mathrm{CMAX}$.
This limit is reached at approximately \SI{15}{\decibel} added path loss for the $-\SI{10}{\decibel}$ setting and \SI{25}{\decibel} for the $+\SI{10}{\decibel}$ setting.
Higher uplink gain therefore reduces the required \gls{ue} transmit power and delays the onset of power-limited operation. 
Unlike the downlink response near its \gls{sinr} ceiling, the uplink response shows a benefit from additional gain through increased transmit-power headroom.

Table~\ref{tab:ran_blueprint} lists the operating point selected from these measurements and used below. The parameter \texttt{ssPBCH\_BlockPower} matches the downlink reference-signal power; changes to downlink gain or attenuation require updating it to avoid biasing the \gls{ue}'s path-loss estimate and open-loop uplink power setting.

\begin{table}[t]
  \centering
  \caption{End-to-end blueprint. The operating point adopted in
  Section~\ref{sec:ran_dynrange} and used for every measurement that follows.}
  \label{tab:ran_blueprint}
  \scriptsize
  \setlength{\tabcolsep}{3pt}
  \begin{tabular}{@{}llc@{}}
    \toprule
    \textbf{Stage} & \textbf{Parameter} & \textbf{Value} \\
    \midrule
    \multirow{2}{*}{\gls{ru} (Foxconn)}
      & Tx attenuation      & \SI{30.0}{\decibel} \\
      & Rx attenuation      & \SI{30.0}{\decibel} \\
    \midrule
    \multirow{2}{*}{Fixed pads}
      & \gls{ru} Tx $\rightarrow$ emu. \gls{gnb} Rx & \SI{20}{\decibel} \\
      & \gls{ue} Tx $\rightarrow$ emu. \gls{ue} Rx  & \SI{20}{\decibel} \\
    \midrule
    \multirow{4}{*}{\lchem}
      & \gls{gnb} port Rx gain & \SI{20}{\decibel} \\
      & \gls{gnb} port Tx gain & \SI{20}{\decibel} \\
      & \gls{ue} port Rx gain  & \SI{30}{\decibel} \\
      & \gls{ue} port Tx gain  & \SI{30}{\decibel} \\
    \midrule
    \multirow{7}{*}{\gls{oai} \gls{du}}
      & \texttt{tx\_amp\_backoff\_dB}     & $30$ \\
      & \texttt{ssPBCH\_BlockPower}        & $-45$ \\
      & \texttt{p0\_NominalWithGrant}      & $-76$ \\
      & \texttt{p0\_nominal}               & $-62$ \\
      & \texttt{preambleRxTargetPower}     & $-60$ \\
      & \texttt{pusch/pucch\_TargetSNRx10} & $270\,/\,130$ \\
      & \texttt{pMax}                      & $6$ \\
    \bottomrule
  \end{tabular}
    \vspace{-.4cm}
\end{table}

\subsection{Downlink and uplink link adaptation}
\label{sec:ran_linkadapt}

\begin{figure}[t]
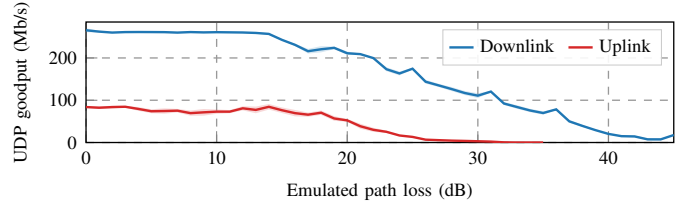

    \centering
    \lchemStack{Figures/ran_baseline_thr}
    \caption{Mean downlink and uplink goodput versus added path loss over ten runs, respectively. Shaded bands indicate 0.95 CI.}
    \label{fig:ran_baseline_thr}
    \vspace{-.3cm}
\end{figure}

\begin{figure}[t]
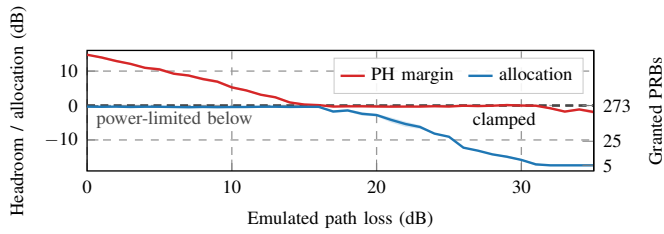

    \centering
    \lchemStack{Figures/ran_baseline_ph_alloc}
    \caption{Uplink power headroom margin and bandwidth
    allocation versus added path loss. 
    Allocation is normalized as 
    $10\log_{10}(N_\mathrm{PRB}/273)$ in dB, 
    where $N_\mathrm{PRB}$ is the number of granted \glspl{prb}.}
    \label{fig:ran_baseline_ph_alloc}
    \vspace{-.3cm}
\end{figure}

Using the configuration in Table~\ref{tab:ran_blueprint}, we sweep the added path loss by scaling the leading \gls{fir} coefficient on all six directional links simultaneously.
We generate bidirectional \gls{udp} traffic with offered loads of \SI[per-mode=symbol]{400}{\mega\bit\per\second} in the downlink and \SI[per-mode=symbol]{150}{\mega\bit\per\second} in the uplink to keep both directions saturated. 
At each path-loss setting, we allow \SI{10}{\second} for the system to settle and then average the measurements over \SI{35}{\second}.
Figs.~\ref{fig:ran_baseline_thr} and~\ref{fig:ran_baseline_ph_alloc} report the results from ten downlink and eleven uplink runs of the same configuration.
We report the mean across runs and 95\% confidence intervals computed using Student's $t$ distribution.

Fig.~\ref{fig:ran_baseline_thr} shows the downlink
and uplink goodput as functions of added path loss.
Downlink goodput decreases from
\SI[per-mode=symbol]{265}{\mega\bit\per\second}
at \SI{0}{\decibel} added path loss to
\SI[per-mode=symbol]{111}{\mega\bit\per\second}
at \SI{30}{\decibel} and
\SI[per-mode=symbol]{70}{\mega\bit\per\second}
at \SI{35}{\decibel}.
It falls below
\SI[per-mode=symbol]{20}{\mega\bit\per\second}
beyond \SI{40}{\decibel}.
Over the same range, the baseline curve in
Fig.~\ref{fig:ran_gain_dl} shows that measured
\gls{rsrp} decreases approximately in proportion
to the programmed attenuation, supporting the
consistency of the applied path-loss sweep.
The 95\% confidence intervals across ten downlink
runs are generally narrow, with larger intervals
in regions where goodput decreases more rapidly.
These results indicate a repeatable mean goodput
response under the selected configuration.

Uplink goodput peaks near
\SI[per-mode=symbol]{85}{\mega\bit\per\second}
and falls to approximately zero at
\SI{33}{\decibel} added path loss, roughly
\SI{10}{\decibel} earlier than the downlink.
Fig.~\ref{fig:ran_baseline_ph_alloc} shows
three operating regimes consistent with a
progressively tighter \gls{ue} transmit-power
constraint.
Up to approximately \SI{16}{\decibel} added
path loss, the scheduler maintains the initial
bandwidth allocation while the power headroom
margin decreases from \SI{14.7}{\decibel}
to approximately zero.
Starting near \SI{17}{\decibel}, the
\gls{phr}-based allocation limit reduces
the grant from 253 to 5 \glspl{prb},
keeping the margin near zero.
Reducing the allocated bandwidth lowers the
total transmit power required to maintain
a given power spectral density.
Beyond approximately \SI{31}{\decibel}
added path loss, the allocation reaches
the configured minimum of five \glspl{prb}.
Further bandwidth reduction is therefore
unavailable, and the power headroom margin
decreases again.

These results show distinct downlink and uplink
responses to increasing added path loss.
Downlink goodput decreases progressively, whereas
the uplink measurements show decreasing
\gls{ue} power headroom followed by reductions
in the scheduled bandwidth.
Under the selected configuration, uplink goodput
falls to approximately zero at an added path
loss roughly \SI{10}{\decibel} lower than
in the downlink.
The experiment demonstrates that \lchem enables
controlled evaluation of these system-level
responses through joint measurements of goodput,
transmit-power headroom, and resource allocation.

\subsection{Multipath Evaluation Using SRS Measurements}
\label{sec:ran_multipath}

Section~\ref{sec:verification} evaluates multipath
emulation using oscilloscope measurements.
Here, we evaluate the channel observed by the
\gls{ran} using the \gls{gnb}'s uplink
\gls{srs} channel estimates.
These measurements capture the end-to-end channel,
including the response programmed into \lchem,
without an external measurement instrument.

\begin{figure}[t]
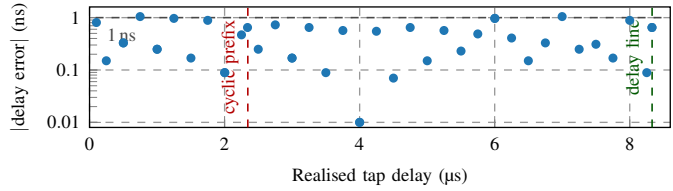

    \centering
    \lchemStack{Figures/ran_srs_twotap}
    \caption{SRS-based delay-estimation error
    relative to realized tap delays for a
    two-tap channel, across 43 captures.}
    \label{fig:ran_srs_twotap}
    \vspace{-.3cm}
\end{figure}

\emph{Two-tap delay sweep.}
We configure a two-tap channel with the first tap
at zero delay and vary the second tap from
\SI{102}{\nano\second} to
\SI{8329}{\nano\second}.
Across 43 \gls{srs} captures, we estimate the
relative tap delay from the recovered \gls{cir}
peaks and compare it with the realized delay
on the emulator's sample grid.
Fig.~\ref{fig:ran_srs_twotap} shows an absolute
error no greater than \SI{1.05}{\nano\second}
over the tested range.
This error characterizes delay recovery relative
to the realized tap positions; it does not
imply programmable delays finer than the
\SI{4.069}{\nano\second} sample spacing.
The sweep includes delays beyond the
\SI{2344}{\nano\second} cyclic prefix.
Recovery of these paths demonstrates their
visibility in the channel estimates, but does
not establish reliable data transmission
at those delays.

\begin{figure}[t]
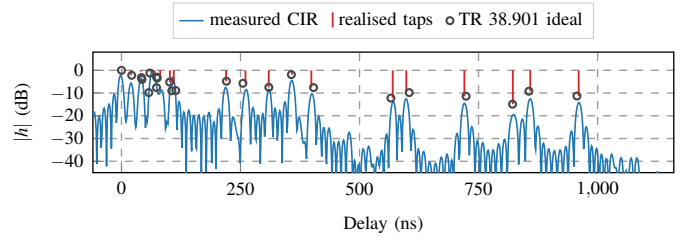

    \centering
    \lchemStack{Figures/ran_srs_tdlb}
    \caption{SRS-based CIR recovery for TDL-B
    with \SI{200}{\nano\second} nominal RMS
    delay spread. Red stems indicate realized
    taps; gray circles indicate requested taps.}
    \label{fig:ran_srs_tdlb}
    \vspace{-.3cm}
\end{figure}

\emph{TDL-B profile.}
We next configure the power-delay profile of
3GPP TR~38.901 TDL-B with a nominal RMS delay
spread of \SI{200}{\nano\second}.
Fig.~\ref{fig:ran_srs_tdlb} compares the
recovered \gls{cir} with the requested and
realized tap profiles.
The measured response follows the realized
profile over its approximately
\SI{1.16}{\micro\second} span.

The reported measurement predates the native-sample MT interface and uses a legacy quantized tap profile.
In that profile, some of the 23 requested taps share a delay, yielding 19 distinct tap positions.
This tap count characterizes the measured configuration rather than the quantization of the current interface.
For each group of coincident taps, we sum
their powers to determine the power of
the merged tap.
This preserves the group's total power,
although delay quantization changes the
original power-delay profile.
We therefore assess the recovered response
against the realized taps.

The displayed \gls{cir} is obtained from a
single \gls{srs} occasion.
We do not average across occasions because
the near-periodic structure of the recovered
response makes their relative delay alignment
ambiguous.
The comparison therefore illustrates recovery
of the configured multipath profile from
one channel estimate.

\section{Conclusion}
\label{sec:con}

The experiments show that a single SDR can provide a programmable propagation medium for both channel-level measurements and an operational 5G RAN.
The sparse filter separates delay span from multiplier cost, while the topology and tap count jointly determine fabric demand.
Fine digital delay and gain settings must be distinguished from measurement accuracy and usable RF dynamic range.
Sample-indexed FPGA phase synthesis removes host-update timing from continuous Doppler generation; interpolation images and measurement floors determine the demonstrated spectral limits. Timestamped snapshots support coordinated parameter changes, while the RAN experiments show why gain calibration and uplink power headroom must be considered when interpreting performance.
These findings guide the selection of channel profiles and operating points for repeatable hardware tests.
The channel-emulation blocks require no custom silicon and no proprietary IP cores, so the approach is reproducible on any X4xx-class radio---a property that matters for the community adoption of hardware-in-the-loop methods.
Future work can extend delay span and link connectivity while evaluating timing closure and Doppler spectral fidelity and snapshot synchronization under the resulting resource and control loads.


\bibliographystyle{IEEEtran}
\bibliography{biblio}

\end{document}

%% file: Figures/tikz_preamble.tex
\usepackage{tikz}
\usepackage{pgfplots}
\pgfplotsset{compat=1.16}
\usepgfplotslibrary{groupplots}
\usepgfplotslibrary{fillbetween}   
\usetikzlibrary{patterns, positioning, fit, arrows.meta}

\definecolor{lchemTheory}{HTML}{484848}
\definecolor{lchemREF}{HTML}{959292}
\definecolor{lchemDL0}{HTML}{1F77B4}
\definecolor{lchemDL1}{HTML}{D62728}
\definecolor{lchemDL2}{HTML}{17A673}
\definecolor{lchemCorrected}{HTML}{A24BFA}
\definecolor{lchemGrid}{HTML}{CCCCCC}     

\definecolor{lchemGainM10}{HTML}{184F95}
\definecolor{lchemGainM5}{HTML}{6DA7EC}
\definecolor{lchemGainBase}{HTML}{52514E}
\definecolor{lchemGainP5}{HTML}{EB6834}
\definecolor{lchemGainP10}{HTML}{B5232F}

\pgfplotsset{colormap={lchemSeq}{rgb255=(242,246,250) rgb255=(143,184,210) rgb255=(35,79,115)}}

\def\lchemPanelWidth{0.35\columnwidth}   
\def\lchemPlotHeight{0.18\columnwidth}   
\def\lchemLegendGap{5pt}                 
\def\lchemWideWidth{0.80\columnwidth}    

\pgfplotsset{
  lchem axis/.style={
    width=0.78\linewidth,
    height=0.56\linewidth,
    scale only axis=true,
    tick align=inside,
    tick pos=left,
    grid=major,
    grid style={dashed, draw=black!40, line width=0.5pt},
    clip mode=individual,
    font=\scriptsize,
    every axis label/.append style={font=\scriptsize},
    tick label style={font=\scriptsize},
    title style={font=\scriptsize},
    legend style={font=\scriptsize},
    lchem legend look,
    axis line style={line width=0.5pt},
    every axis plot/.append style={line width=0.8pt},
  },
  lchem legend look/.style={
    legend cell align=left,
    legend columns=-1,
    legend image code/.code={%
      \draw[mark repeat=2, mark phase=2, ##1]
        plot coordinates {(0cm,0cm) (0.16cm,0cm) (0.32cm,0cm)};},
    legend style={
      font=\scriptsize, draw=lchemGrid, fill=white, fill opacity=0.9,
      text opacity=1, inner xsep=2pt, inner ysep=1.5pt,
      /tikz/every even column/.append style={column sep=3pt},
    },
  },
  lchem panel/.style={lchem axis, width=\lchemPanelWidth, height=\lchemPlotHeight},
  lchem wide/.style={lchem axis, width=\lchemWideWidth, height=\lchemPlotHeight},
  lchem top legend/.style={legend style={at={(0.5,1)}, anchor=south,
    yshift=\lchemLegendGap}},
  lchem shared legend/.style={
    hide axis,
    width=0.95\columnwidth, height=1pt,
    scale only axis=true,
    xmin=0, xmax=1, ymin=0, ymax=1,
    lchem legend look,
    legend style={at={(0.5,0.5)}, anchor=center},
  },
  lchem dl0/.style={color=lchemDL0, mark=o, mark size=2.4pt,
    mark options={draw=lchemDL0, line width=0.5pt}},
  lchem dl1/.style={color=lchemDL1, mark=square*, mark size=2.4pt,
    mark options={fill=lchemDL1!30, draw=lchemDL1, line width=0.5pt}},
  lchem dl2/.style={color=lchemDL2, mark=triangle*, mark size=2.6pt,
    mark options={fill=lchemDL2!30, draw=lchemDL2, line width=0.5pt}},
  lchem measured/.style={color=lchemDL0, mark=*, mark size=1.8pt,
    mark options={fill=lchemDL0, draw=lchemDL0, line width=0.4pt}},
  lchem trace/.style={color=lchemDL0, line width=0.7pt, no marks},
  lchem stem legend/.style={legend image code/.code={%
    \draw[##1] (0.16cm,-0.09cm) -- (0.16cm,0.11cm);}},
  lchem tap/.style={only marks, mark=x, mark size=4pt, color=lchemTheory,
    mark options={line width=1.2pt}},
  lchem ideal/.style={
    densely dashed,
    color=lchemTheory,
    line width=1.0pt,
  },
}

\tikzset{
  lchem ideal/.style={
    densely dashed,
    color=lchemTheory,
    line width=1.0pt,
  },
}

\newdimen\lchemOverL   
\newdimen\lchemOverR   
\newdimen\lchemAxisW   
\newdimen\lchemOverT   
\tikzset{lchem measure/.style={execute at end picture={%
  \pgfextractx{\lchemOverL}{\pgfpointdiff
    {\pgfpointanchor{current bounding box}{south west}}
    {\pgfpointanchor{current axis}{south west}}}%
  \pgfextractx{\lchemAxisW}{\pgfpointdiff
    {\pgfpointanchor{current axis}{south west}}
    {\pgfpointanchor{current axis}{south east}}}%
  \pgfextractx{\lchemOverR}{\pgfpointdiff
    {\pgfpointanchor{current axis}{south east}}
    {\pgfpointanchor{current bounding box}{north east}}}%
  \pgfextracty{\lchemOverT}{\pgfpointdiff
    {\pgfpointanchor{current axis}{north east}}
    {\pgfpointanchor{current bounding box}{north east}}}%
  \global\lchemOverT=\lchemOverT
  \global\lchemOverL=\lchemOverL
  \global\lchemOverR=\lchemOverR
  \global\lchemAxisW=\lchemAxisW}}}

\newsavebox\lchemBoxA
\newsavebox\lchemBoxB
\newlength\lchemAL \newlength\lchemAW \newlength\lchemAR
\newlength\lchemBL \newlength\lchemBW \newlength\lchemBR
\newlength\lchemCA \newlength\lchemCB \newlength\lchemCapW
\newlength\lchemShort \newlength\lchemMoveA \newlength\lchemMoveB
\newlength\lchemAT \newlength\lchemBT \newlength\lchemRise
\def\lchemPanelGap{4pt}   

\newcommand\lchemMinTo[2]{\ifdim#1>\dimexpr#2\relax\setlength#1{\dimexpr#2\relax}\fi}
\newcommand\lchemMaxTo[2]{\ifdim#1<\dimexpr#2\relax\setlength#1{\dimexpr#2\relax}\fi}

\newif\iflchemAfterLegend

\newcommand\lchemPanel[5]{%
  \begin{subfigure}[t]{\lchemCapW}%
    \hbox to \lchemCapW{\kern\dimexpr0.5\lchemCapW-0.5#3-#2\relax\usebox#1\hss}%
    \caption{#4}\label{#5}%
  \end{subfigure}}

\newcommand\lchemPanelRow[6]{%
  \begingroup
  \sbox\lchemBoxA{\tikzset{every picture/.append style={lchem measure}}\input{#1}}%
  \setlength\lchemAL{\lchemOverL}\setlength\lchemAW{\lchemAxisW}\setlength\lchemAR{\lchemOverR}\setlength\lchemAT{\lchemOverT}%
  \sbox\lchemBoxB{\tikzset{every picture/.append style={lchem measure}}\input{\lchemBW}}%
  \setlength\lchemBL{\lchemOverL}\setlength\lchemBW{\lchemAxisW}\setlength\lchemBR{\lchemOverR}\setlength\lchemBT{\lchemOverT}%
  \setlength\lchemShort{\dimexpr\lchemAL+\lchemAW+\lchemAR+\lchemPanelGap
    +\lchemBL+\lchemBW+\lchemBR-\linewidth\relax}%
  \ifdim\lchemShort>0pt
    \edef\lchemPanelWidth{\the\dimexpr\lchemAW-0.5\lchemShort-0.5pt\relax}%
  \sbox\lchemBoxA{\tikzset{every picture/.append style={lchem measure}}\input{#1}}%
  \setlength\lchemAL{\lchemOverL}\setlength\lchemAW{\lchemAxisW}\setlength\lchemAR{\lchemOverR}
  \sbox\lchemBoxB{\tikzset{every picture/.append style={lchem measure}}\input{\lchemBW}}%
  \setlength\lchemBL{\lchemOverL}\setlength\lchemBW{\lchemAxisW}\setlength\lchemBR{\lchemOverR}
  \fi
  \setlength\lchemCA{0.25\linewidth}%
  \lchemMaxTo\lchemCA{\lchemAL+0.5\lchemAW}%
  \setlength\lchemCB{0.75\linewidth}%
  \lchemMinTo\lchemCB{\linewidth-\lchemBR-0.5\lchemBW}%
  \setlength\lchemShort{\dimexpr0.5\lchemAW+\lchemAR+\lchemPanelGap
    +\lchemBL+0.5\lchemBW-\lchemCB+\lchemCA\relax}%
  \ifdim\lchemShort>0pt
    \setlength\lchemMoveA{0.5\lchemShort}%
    \lchemMinTo\lchemMoveA{\lchemCA-\lchemAL-0.5\lchemAW}%
    \setlength\lchemMoveB{\dimexpr\lchemShort-\lchemMoveA\relax}%
    \lchemMinTo\lchemMoveB{\linewidth-\lchemBR-0.5\lchemBW-\lchemCB}%
    \setlength\lchemMoveA{\dimexpr\lchemShort-\lchemMoveB\relax}%
    \lchemMinTo\lchemMoveA{\lchemCA-\lchemAL-0.5\lchemAW}%
    \addtolength\lchemCA{-\lchemMoveA}%
    \addtolength\lchemCB{\lchemMoveB}%
  \fi
  \setlength\lchemCapW{\dimexpr\lchemCB-\lchemCA\relax}%
  \lchemMinTo\lchemCapW{2\lchemCA}%
  \lchemMinTo\lchemCapW{2\linewidth-2\lchemCB}%
  \par
  \iflchemAfterLegend
    \global\lchemAfterLegendfalse
    \setlength\lchemRise{\dimexpr\ht\lchemBoxA-\lchemAT\relax}%
    \lchemMaxTo\lchemRise{\ht\lchemBoxB-\lchemBT}%
    \setlength\lchemShort{\ht\lchemBoxA}\lchemMaxTo\lchemShort{\ht\lchemBoxB}%
    \addtolength\lchemRise{-\lchemShort}%
    \vspace*{\dimexpr\lchemLegendGap+\lchemRise-\lineskip\relax}%
  \fi
  \noindent\makebox[\linewidth][l]{%
    \hspace*{\dimexpr\lchemCA-0.5\lchemCapW\relax}%
    \lchemPanel\lchemBoxA\lchemAL\lchemAW{#2}{#3}%
    \hspace*{\dimexpr\lchemCB-\lchemCA-\lchemCapW\relax}%
    \lchemPanel\lchemBoxB\lchemBL\lchemBW{#5}{#6}%
    \hss}\par
  \endgroup}

\makeatletter
\newcommand\lchemStack[1]{%
  \setlength\lchemBL{0pt}\setlength\lchemBR{0pt}%
  \@for\lchem@file:=#1\do{%
  \sbox\lchemBoxA{\tikzset{every picture/.append style={lchem measure}}\input{\lchem@file}}%
  \setlength\lchemAL{\lchemOverL}\setlength\lchemAW{\lchemAxisW}\setlength\lchemAR{\lchemOverR}
    \lchemMaxTo\lchemBL{\lchemAL}\lchemMaxTo\lchemBR{\lchemAR}}%
  \begingroup
  \edef\lchemWideWidth{\the\dimexpr\linewidth-\lchemBL-\lchemBR\relax}%
  \@for\lchem@file:=#1\do{%
  \sbox\lchemBoxA{\tikzset{every picture/.append style={lchem measure}}\input{\lchem@file}}%
  \setlength\lchemAL{\lchemOverL}\setlength\lchemAW{\lchemAxisW}\setlength\lchemAR{\lchemOverR}
    \par\noindent\makebox[\linewidth][l]{%
      \kern\dimexpr\lchemBL-\lchemAL\relax\usebox\lchemBoxA\hss}\par}%
  \endgroup}
\makeatother

%% file: acronym.tex
\newacronym{3gpp}{3GPP}{3rd Generation Partnership Project}
\newacronym{4g}{4G}{4th generation mobile network}
\newacronym{5g}{5G}{5th generation mobile network}
\newacronym{6g}{6G}{6th generation mobile network}
\newacronym{nextg}{NextG}{Next Generation}
\newacronym{5gc}{5GC}{5G Core}
\newacronym{adc}{ADC}{Analog to Digital Converter}
\newacronym{dac}{DAC}{Digital to Analog Converter}
\newacronym{aerpaw}{AERPAW}{Aerial Experimentation and Research Platform for Advanced Wireless}
\newacronym{ai}{AI}{Artificial Intelligence}
\newacronym{aimd}{AIMD}{Additive Increase Multiplicative Decrease}
\newacronym{am}{AM}{Acknowledged Mode}
\newacronym{amc}{AMC}{Adaptive Modulation and Coding}
\newacronym{amf}{AMF}{Access and Mobility Management Function}
\newacronym{aops}{AOPS}{Adaptive Order Prediction Scheduling}
\newacronym{api}{API}{Application Programming Interface}
\newacronym{apn}{APN}{Access Point Name}
\newacronym{aqm}{AQM}{Active Queue Management}
\newacronym{ausf}{AUSF}{Authentication Server Function}
\newacronym{avc}{AVC}{Advanced Video Coding}
\newacronym{awgn}{AGWN}{Additive White Gaussian Noise}
\newacronym{balia}{BALIA}{Balanced Link Adaptation Algorithm}
\newacronym{bbu}{BBU}{Base Band Unit}
\newacronym{bdp}{BDP}{Bandwidth-Delay Product}
\newacronym{ber}{BER}{Bit Error Rate}
\newacronym{bf}{BF}{Beamforming}
\newacronym{bler}{BLER}{Block Error Rate}
\newacronym{brr}{BRR}{Bayesian Ridge Regressor}
\newacronym{bsr}{BSR}{Buffer Status Report}
\newacronym{bs}{BS}{Base Station}
\newacronym{bpsk}{BPSK}{Binary Phase-shift keying}
\newacronym{bss}{BSS}{Business Support System}
\newacronym{ca}{CA}{Carrier Aggregation}
\newacronym{caas}{CaaS}{Connectivity-as-a-Service}
\newacronym{cb}{CB}{Code Block}
\newacronym{cc}{CC}{Congestion Control}
\newacronym{ccid}{CCID}{Congestion Control ID}
\newacronym{cco}{CC}{Carrier Component}
\newacronym{cd}{CD}{Continuous Delivery}
\newacronym{cdd}{CDD}{Cyclic Delay Diversity}
\newacronym{cdf}{CDF}{Cumulative Distribution Function}
\newacronym{cdma}{CDMA}{Code-Division Multiple Access}
\newacronym{cdn}{CDN}{Content Distribution Network}
\newacronym{ci}{CI}{Continuous Integration}
\newacronym{cicd}{CI/CD}{Continuous Integration/Continuous Delivery}
\newacronym{cir}{CIR}{Channel Impulse Response}
\newacronym{cfr}{CFR}{Channel Frequency Response}
\newacronym{cn}{CN}{Core Network}
\newacronym{codel}{CoDel}{Controlled Delay Management}
\newacronym{comac}{COMAC}{Converged Multi-Access and Core}
\newacronym{cord}{CORD}{Central Office Re-architected as a Datacenter}
\newacronym{cornet}{CORNET}{COgnitive Radio NETwork}
\newacronym{cosmos}{COSMOS}{Cloud Enhanced Open Software Defined Mobile Wireless Testbed for City-Scale Deployment}
\newacronym{cots}{COTS}{Commercial Off-the-Shelf}
\newacronym{cp}{CP}{Control Plane}
\newacronym{cpu}{CPU}{Central Processing Unit}
\newacronym{cqi}{CQI}{Channel Quality Information}
\newacronym{cr}{CR}{Cognitive Radio}
\newacronym{cran}{CRAN}{Cloud \gls{ran}}
\newacronym{crs}{CRS}{Cell Reference Signal}
\newacronym{csi}{CSI}{Channel State Information}
\newacronym{csirs}{CSI-RS}{Channel State Information - Reference Signal}
\newacronym{cu}{CU}{Central Unit}
\newacronym{d2tcp}{D$^2$TCP}{Deadline-aware Data center TCP}
\newacronym{d3}{D$^3$}{Deadline-Driven Delivery}
\newacronym{dag}{DAG}{Directed Acyclic Graph}
\newacronym{darpa}{DARPA}{Defense Advanced Research Projects Agency}
\newacronym{das}{DAS}{Distributed Antenna System}
\newacronym{dash}{DASH}{Dynamic Adaptive Streaming over HTTP}
\newacronym{dbs}{DBS}{Deep Brain Stimulation}
\newacronym{dc}{DC}{Dual Connectivity}
\newacronym{dccp}{DCCP}{Datagram Congestion Control Protocol}
\newacronym{dce}{DCE}{Direct Code Execution}
\newacronym{dci}{DCI}{Downlink Control Information}
\newacronym{dcl}{DCL}{Dear Colleague Letter}
\newacronym{dctcp}{DCTCP}{Data Center TCP}
\newacronym{devops}{DevOps}{Development and Operations}
\newacronym{dl}{DL}{Downlink}
\newacronym{dmr}{DMR}{Deadline Miss Ratio}
\newacronym{dmrs}{DMRS}{DeModulation Reference Signal}
\newacronym{drlcc}{DRL-CC}{Deep Reinforcement Learning Congestion Control}
\newacronym{drs}{DRS}{Discovery Reference Signal}
\newacronym{dt}{DT}{Digital Twin}
\newacronym{dtn}{DTN}{Digital Twin Network}
\newacronym{dtmn}{DTMN}{Digital Twin for Mobile Network}
\newacronym{dtwn}{DTWN}{Digital Twin Wireless Network}
\newacronym{du}{DU}{Distributed Unit}
\newacronym{e2e}{E2E}{end-to-end}
\newacronym{ecaas}{ECaaS}{Edge-Cloud-as-a-Service}
\newacronym{ecn}{ECN}{Explicit Congestion Notification}
\newacronym{edf}{EDF}{Earliest Deadline First}
\newacronym{em}{EM}{Electro-Magnetic}
\newacronym{embb}{eMBB}{Enhanced Mobile Broadband}
\newacronym{empower}{EMPOWER}{EMpowering transatlantic PlatfOrms for advanced WirEless Research}
\newacronym{enb}{eNB}{evolved Node Base}
\newacronym{endc}{EN-DC}{E-UTRAN-\gls{nr} \gls{dc}}
\newacronym{epc}{EPC}{Evolved Packet Core}
\newacronym{eps}{EPS}{Evolved Packet System}
\newacronym{evm}{EVM}{Error Vector Magnitude}
\newacronym{es}{ES}{Edge Server}
\newacronym{etsi}{ETSI}{European Telecommunications Standards Institute}
\newacronym[firstplural=Estimated Times of Arrival (ETAs)]{eta}{ETA}{Estimated Time of Arrival}
\newacronym{eutran}{E-UTRAN}{Evolved Universal Terrestrial Access Network}
\newacronym{faas}{FaaS}{Function-as-a-Service}
\newacronym{fapi}{FAPI}{Functional Application Platform Interface}
\newacronym{fcc}{FCC}{Federal Communications Commission}
\newacronym{fdd}{FDD}{Frequency Division Duplexing}
\newacronym{fdm}{FDM}{Frequency Division Multiplexing}
\newacronym{fdma}{FDMA}{Frequency Division Multiple Access}
\newacronym{fed4fire}{FED4FIRE+}{Federation 4 Future Internet Research and Experimentation Plus}
\newacronym{fir}{FIR}{Finite Impulse Response}
\newacronym{fit}{FIT}{Future \acrlong{iot}}
\newacronym{fl}{FL}{Federated Learning}
\newacronym{fpga}{FPGA}{Field Programmable Gate Array}
\newacronym{fr2}{FR2}{Frequency Range 2}
\newacronym{fs}{FS}{Fast Switching}
\newacronym{fscc}{FSCC}{Flow Sharing Congestion Control}
\newacronym{ftp}{FTP}{File Transfer Protocol}
\newacronym{fw}{FW}{Flow Window}
\newacronym{ga128}{Ga}{Golay Sequence type A}
\newacronym{ge}{GE}{Gaussian Elimination}
\newacronym{glfsr}{GLFSR}{Galois Linear Feedback Shift Register}
\newacronym{gnb}{gNB}{Next Generation Node Base}
\newacronym{gold}{Gold}{Gold}
\newacronym{gop}{GOP}{Group of Pictures}
\newacronym{gpr}{GPR}{Gaussian Process Regressor}
\newacronym{gpu}{GPU}{Graphics Processing Unit}
\newacronym{gtp}{GTP}{GPRS Tunneling Protocol}
\newacronym{gtpc}{GTP-C}{GPRS Tunnelling Protocol Control Plane}
\newacronym{gtpu}{GTP-U}{GPRS Tunnelling Protocol User Plane}
\newacronym{gtpv2c}{GTPv2-C}{\gls{gtp} v2 - Control}
\newacronym{gw}{GW}{Gateway}
\newacronym{harq}{HARQ}{Hybrid Automatic Repeat reQuest}
\newacronym{hetnet}{HetNet}{Heterogeneous Network}
\newacronym{hh}{HH}{Hard Handover}
\newacronym{hol}{HOL}{Head-of-Line}
\newacronym{hqf}{HQF}{Highest-quality-first}
\newacronym{hss}{HSS}{Home Subscription Server}
\newacronym{http}{HTTP}{HyperText Transfer Protocol}
\newacronym{ia}{IA}{Initial Access}
\newacronym{iab}{IAB}{Integrated Access and Backhaul}
\newacronym{ic}{IC}{Incident Command}
\newacronym{ietf}{IETF}{Internet Engineering Task Force}
\newacronym{ifw}{IFW}{Interference Free Window}
\newacronym{imsi}{IMSI}{International Mobile Subscriber Identity}
\newacronym{imt}{IMT}{International Mobile Telecommunication}
\newacronym{iot}{IoT}{Internet of Things}
\newacronym{ip}{IP}{Internet Protocol}
\newacronym{iq}{IQ}{In-phase and Quadrature}
\newacronym{isi}{ISI}{Inter-Symbol Interference}
\newacronym{itu}{ITU}{International Telecommunication Union}
\newacronym{kpi}{KPI}{Key Performance Indicator}
\newacronym{kvm}{KVM}{Kernel-based Virtual Machine}
\newacronym{lfsr}{LFSR}{Linear Feedback Shift Register}
\newacronym{los}{LOS}{Line-of-Sight}
\newacronym{ls}{LS}{Loosely Synchronised}
\newacronym{lsm}{LSM}{Link-to-System Mapping}
\newacronym{lstm}{LSTM}{Long Short Term Memory}
\newacronym{lte}{LTE}{Long Term Evolution}
\newacronym{lxc}{LXC}{Linux Container}
\newacronym{m2m}{M2M}{Machine to Machine}
\newacronym{mac}{MAC}{Medium Access Control}
\newacronym{mai}{MAI}{Multiple Access Interference}
\newacronym{manet}{MANET}{Mobile Ad Hoc Network}
\newacronym{mano}{MANO}{Management and Orchestration}
\newacronym{mc}{MC}{Multi-Connectivity}
\newacronym{mcc}{MCC}{Mobile Cloud Computing}
\newacronym{mchem}{MCHEM}{Massive Channel Emulator}
\newacronym{mcs}{MCS}{Modulation and Coding Scheme}
\newacronym{mec}{MEC}{Multi-access Edge Computing}
\newacronym{mec2}{MEC}{Mobile Edge Cloud}
\newacronym{mec3}{MEC}{Mobile Edge Computing}
\newacronym{mfc}{MFC}{Mobile Fog Computing}
\newacronym{mi}{MI}{Mutual Information}
\newacronym{mib}{MIB}{Master Information Block}
\newacronym{miesm}{MIESM}{Mutual Information Based Effective SINR}
\newacronym{mimo}{MIMO}{Multiple Input, Multiple Output}
\newacronym{mgen}{MGEN}{Multi-Generator}
\newacronym{ml}{ML}{Machine Learning}
\newacronym{mlr}{MLR}{Maximum-local-rate}
\newacronym[plural=\gls{mme}s,firstplural=Mobility Management Entities (MMEs)]{mme}{MME}{Mobility Management Entity}
\newacronym{mmtc}{mMTC}{Massive Machine-Type Communications}
\newacronym{mmwave}{mmWave}{millimeter wave}
\newacronym{mpdccp}{MP-DCCP}{Multipath Datagram Congestion Control Protocol}
\newacronym{mptcp}{MPTCP}{Multipath TCP}
\newacronym{mr}{MR}{Maximum Rate}
\newacronym{mrdc}{MR-DC}{Multi \gls{rat} \gls{dc}}
\newacronym{mse}{MSE}{Mean Square Error}
\newacronym{mss}{MSS}{Maximum Segment Size}
\newacronym{mt}{MT}{Mobile Termination}
\newacronym{mtd}{MTD}{Machine-Type Device}
\newacronym{mtu}{MTU}{Maximum Transmission Unit}
\newacronym{mumimo}{MU-MIMO}{Multi-user \gls{mimo}}
\newacronym{mvno}{MVNO}{Mobile Virtual Network Operator}
\newacronym{nalu}{NALU}{Network Abstraction Layer Unit}
\newacronym{nas}{NAS}{Network Attached Storage}
\newacronym{nbiot}{NB-IoT}{Narrow Band IoT}
\newacronym{nfv}{NFV}{Network Function Virtualization}
\newacronym{nfvi}{NFVI}{Network Function Virtualization Infrastructure}
\newacronym{nic}{NIC}{Network Interface Card}
\newacronym{nlos}{NLOS}{Non-Line-of-Sight}
\newacronym{now}{NOW}{Non Overlapping Window}
\newacronym{nrdz}{NRDZ}{National Radio Dynamic Zone}
\newacronym{nco}{NCO}{Numerically Controlled Oscillator}
\newacronym{nsf}{NSF}{National Science Foundation}
\newacronym{nsm}{NSM}{Network Service Mesh}
\newacronym[type=hidden]{nr}{NR}{New Radio}
\newacronym{nrf}{NRF}{Network Repository Function}
\newacronym{nsa}{NSA}{Non Stand Alone}
\newacronym{nse}{NSE}{Network Slicing Engine}
\newacronym{nssf}{NSSF}{Network Slice Selection Function}
\newacronym{ntp}{NTP}{Network Time Protocol}
\newacronym{o2i}{O2I}{Outdoor to Indoor}
\newacronym{oai}{OAI}{OpenAirInterface}
\newacronym{oaicn}{OAI-CN}{\gls{oai} \acrlong{cn}}
\newacronym{oairan}{OAI-RAN}{\acrlong{oai} \acrlong{ran}}
\newacronym{oam}{OAM}{Operations, Administration and Maintenance}
\newacronym[plural=\gls{obu}s,firstplural=Onboard Units (OBUs)]{obu}{OBU}{Onboard Unit}
\newacronym{ofdm}{OFDM}{Orthogonal Frequency Division Multiplexing}
\newacronym{olia}{OLIA}{Opportunistic Linked Increase Algorithm}
\newacronym{omec}{OMEC}{Open Mobile Evolved Core}
\newacronym{onap}{ONAP}{Open Network Automation Platform}
\newacronym{onf}{ONF}{Open Networking Foundation}
\newacronym{onos}{ONOS}{Open Networking Operating System}
\newacronym{oom}{OOM}{\gls{onap} Operations Manager}
\newacronym{opnfv}{OPNFV}{Open Platform for \gls{nfv}}
\newacronym[type=hidden]{oran}{O-RAN}{Open RAN}
\newacronym{orbit}{ORBIT}{Open-Access Research Testbed for Next-Generation Wireless Networks}
\newacronym{os}{OS}{Operating System}
\newacronym{osm}{OSM}{Open Street Map}
\newacronym{oss}{OSS}{Operations Support System}
\newacronym{pa}{PA}{Position-aware}
\newacronym{pase}{PASE}{Prioritization, Arbitration, and Self-adjusting Endpoints}
\newacronym{pawr}{PAWR}{Platforms for Advanced Wireless Research}
\newacronym{pbch}{PBCH}{Physical Broadcast Channel}
\newacronym{pcef}{PCEF}{Policy and Charging Enforcement Function}
\newacronym{pcfich}{PCFICH}{Physical Control Format Indicator Channel}
\newacronym{pcrf}{PCRF}{Policy and Charging Rules Function}
\newacronym{pdcch}{PDCCH}{Physical Downlink Control Channel}
\newacronym{pdcp}{PDCP}{Packet Data Convergence Protocol}
\newacronym{pdsch}{PDSCH}{Physical Downlink Shared Channel}
\newacronym{pdu}{PDU}{Packet Data Unit}
\newacronym{pdp}{PDP}{Power Delay Profile}
\newacronym{pf}{PF}{Proportional Fair}
\newacronym{pgw}{PGW}{Packet Gateway}
\newacronym{phich}{PHICH}{Physical Hybrid ARQ Indicator Channel}
\newacronym{phy}{PHY}{Physical}
\newacronym{pl}{PL}{Path Loss}
\newacronym{pmch}{PMCH}{Physical Multicast Channel}
\newacronym{pmi}{PMI}{Precoding Matrix Indicators}
\newacronym{powder}{POWDER}{Platform for Open Wireless Data-driven Experimental Research}
\newacronym{ppo}{PPO}{Proximal Policy Optimization}
\newacronym{ppp}{PPP}{Poisson Point Process}
\newacronym{prach}{PRACH}{Physical Random Access Channel}
\newacronym{prb}{PRB}{Physical Resource Block}
\newacronym{psnr}{PSNR}{Peak Signal to Noise Ratio}
\newacronym{pss}{PSS}{Primary Synchronization Signal}
\newacronym{pucch}{PUCCH}{Physical Uplink Control Channel}
\newacronym{pusch}{PUSCH}{Physical Uplink Shared Channel}
\newacronym{qam}{QAM}{Quadrature Amplitude Modulation}
\newacronym{qci}{QCI}{\gls{qos} Class Identifier}
\newacronym{qoe}{QoE}{Quality of Experience}
\newacronym{qos}{QoS}{Quality of Service}
\newacronym{qtgui}{QT-GUI}{QT Graphical User Interface}
\newacronym{quic}{QUIC}{Quick UDP Internet Connections}
\newacronym{rach}{RACH}{Random Access Channel}
\newacronym{ran}{RAN}{Radio Access Network}
\newacronym[firstplural=Radio Access Technologies (RATs)]{rat}{RAT}{Radio Access Technology}
\newacronym{rcn}{RCN}{Research Coordination Network}
\newacronym{rec}{REC}{Radio Edge Cloud}
\newacronym{red}{RED}{Random Early Detection}
\newacronym{renew}{RENEW}{Reconfigurable Eco-system for Next-generation End-to-end Wireless}
\newacronym{rf}{RF}{Radio Frequency}
\newacronym{rfc}{RFC}{Request for Comments}
\newacronym{rfr}{RFR}{Random Forest Regressor}
\newacronym{ric}{RIC}{RAN Intelligent Controller}
\newacronym{rlc}{RLC}{Radio Link Control}
\newacronym{rlf}{RLF}{Radio Link Failure}
\newacronym{rlnc}{RLNC}{Random Linear Network Coding}
\newacronym{rmse}{RMSE}{Root Mean Squared Error}
\newacronym{rnis}{RNIS}{Radio Network Information Service}
\newacronym{rr}{RR}{Round Robin}
\newacronym{rrc}{RRC}{Radio Resource Control}
\newacronym{rrm}{RRM}{Radio Resource Management}
\newacronym{rru}{RRU}{Remote Radio Unit}
\newacronym{rs}{RS}{Remote Server}
\newacronym{rsrp}{RSRP}{Reference Signal Received Power}
\newacronym{rsrq}{RSRQ}{Reference Signal Received Quality}
\newacronym{rss}{RSS}{Received Signal Strength}
\newacronym{rssi}{RSSI}{Received Signal Strength Indicator}
\newacronym{rsu}{RSU}{Road-Side Unit}
\newacronym{rtt}{RTT}{Round Trip Time}
\newacronym{ru}{RU}{Radio Unit}
\newacronym{rw}{RW}{Receive Window}
\newacronym{rx}{RX}{Receiver}
\newacronym{s1ap}{S1AP}{S1 Application Protocol}
\newacronym{sa}{SA}{standalone}
\newacronym{sack}{SACK}{Selective Acknowledgment}
\newacronym{sap}{SAP}{Service Access Point}
\newacronym{sc2}{SC2}{Spectrum Collaboration Challenge}
\newacronym{scef}{SCEF}{Service Capability Exposure Function}
\newacronym{sch}{SCH}{Secondary Cell Handover}
\newacronym{scoot}{SCOOT}{Split Cycle Offset Optimization Technique}
\newacronym{sctp}{SCTP}{Stream Control Transmission Protocol}
\newacronym{sdap}{SDAP}{Service Data Adaptation Protocol}
\newacronym{sd}{SD}{Standard Deviation}
\newacronym{sdk}{SDK}{Software Development Kit}
\newacronym{sdm}{SDM}{Space Division Multiplexing}
\newacronym{sdma}{SDMA}{Spatial Division Multiple Access}
\newacronym{sdn}{SDN}{Software-defined Networking}
\newacronym{sdr}{SDR}{Software-defined Radio}
\newacronym{seba}{SEBA}{SDN-Enabled Broadband Access}
\newacronym{sgsn}{SGSN}{Serving GPRS Support Node}
\newacronym{sgw}{SGW}{Service Gateway}
\newacronym{si}{SI}{Study Item}
\newacronym{sib}{SIB}{Secondary Information Block}
\newacronym{sinr}{SINR}{Signal to Interference plus Noise Ratio}
\newacronym{sip}{SIP}{Session Initiation Protocol}
\newacronym{siso}{SISO}{Single Input, Single Output}
\newacronym{sla}{SLA}{Service Level Agreement}
\newacronym{sm}{SM}{Saturation Mode}
\newacronym{smf}{SMF}{Session Management Function}
\newacronym{smo}{SMO}{Service Management and Orchestration}
\newacronym{sms}{SMS}{Short Message Service}
\newacronym{smsgmsc}{SMS-GMSC}{\gls{sms}-Gateway}
\newacronym{snr}{SNR}{Signal-to-Noise-Ratio}
\newacronym{son}{SON}{Self-Organizing Network}
\newacronym{sptcp}{SPTCP}{Single Path TCP}
\newacronym{srb}{SRB}{Service Radio Bearer}
\newacronym{srn}{SRN}{Standard Radio Node}
\newacronym{srs}{SRS}{Sounding Reference Signal}
\newacronym{ss}{SS}{Synchronization Signal}
\newacronym{sss}{SSS}{Secondary Synchronization Signal}
\newacronym{st}{ST}{Spanning Tree}
\newacronym{svc}{SVC}{Scalable Video Coding}
\newacronym{tb}{TB}{Transport Block}
\newacronym{tcp}{TCP}{Transmission Control Protocol}
\newacronym{tdd}{TDD}{Time Division Duplexing}
\newacronym{tpc}{TPC}{Transmit Power Control}
\newacronym{phr}{PHR}{Power Headroom Report}
\newacronym{tdm}{TDM}{Time Division Multiplexing}
\newacronym{tdma}{TDMA}{Time Division Multiple Access}
\newacronym{tfl}{TfL}{Transport for London}
\newacronym{tfrc}{TFRC}{TCP-Friendly Rate Control}
\newacronym{tft}{TFT}{Traffic Flow Template}
\newacronym{tgen}{TGEN}{Traffic Generator}
\newacronym{tip}{TIP}{Telecom Infra Project}
\newacronym{tm}{TM}{Transparent Mode}
\newacronym{to}{TO}{Telco Operator}
\newacronym{toa}{ToA}{Time of Arrival}
\newacronym{tr}{TR}{Technical Report}
\newacronym{trp}{TRP}{Transmitter Receiver Pair}
\newacronym{ts}{TS}{Technical Specification}
\newacronym{tti}{TTI}{Transmission Time Interval}
\newacronym{ttt}{TTT}{Time-to-Trigger}
\newacronym{tx}{TX}{Transmitter}
\newacronym{uas}{UAS}{Unmanned Aerial System}
\newacronym{uav}{UAV}{Unmanned Aerial Vehicle}
\newacronym{udm}{UDM}{Unified Data Management}
\newacronym{udp}{UDP}{User Datagram Protocol}
\newacronym{udr}{UDR}{Unified Data Repository}
\newacronym{ue}{UE}{User Equipment}
\newacronym{uhd}{UHD}{\gls{usrp} Hardware Driver}
\newacronym{ul}{UL}{Uplink}
\newacronym{um}{UM}{Unacknowledged Mode}
\newacronym{uml}{UML}{Unified Modeling Language}
\newacronym{upa}{UPA}{Uniform Planar Array}
\newacronym{upf}{UPF}{User Plane Function}
\newacronym{urllc}{URLLC}{Ultra Reliable and Low Latency Communication}
\newacronym{usa}{U.S.}{United States}
\newacronym{usim}{USIM}{Universal Subscriber Identity Module}
\newacronym{usrp}{USRP}{Universal Software Radio Peripheral}
\newacronym{utc}{UTC}{Urban Traffic Control}
\newacronym{vim}{VIM}{Virtualization Infrastructure Manager}
\newacronym{vm}{VM}{Virtual Machine}
\newacronym{vnf}{VNF}{Virtual Network Function}
\newacronym{volte}{VoLTE}{Voice over \gls{lte}}
\newacronym{voltha}{VOLTHA}{Virtual OLT HArdware Abstraction}
\newacronym{vr}{VR}{Virtual Reality}
\newacronym{vran}{vRAN}{Virtualized \gls{ran}}
\newacronym{vss}{VSS}{Video Streaming Server}
\newacronym{wbf}{WBF}{Wired Bias Function}
\newacronym{wf}{WF}{Wired-first}
\newacronym{wi}{WI}{Wireless InSite}
\newacronym{wlan}{WLAN}{Wireless Local Area Network}
\newacronym{pnf}{PNF}{Physical Network Function}
\newacronym{drl}{DRL}{Deep Reinforcement Learning}
\newacronym{mtc}{MTC}{Machine-type Communications}
\newacronym{v2x}{V2X}{Vehicle-to-everything}
\newacronym{cast}{CaST}{Channel emulation scenario generator and Sounder Toolchain}
\newacronym{gui}{GUI}{Graphical User Interface}
\newacronym{ups}{UPS}{Uninterruptible Power Supply}
\newacronym{ota}{OTA}{Over-the-Air}
\newacronym{hitl}{HITL}{hardware-in-the-loop}
\newacronym{soc}{SoC}{System-on-Chip}
\newacronym{eeg}{EEG}{electroencephalogram}
\newacronym{ieeg}{iEEG}{intracranial electroencephalogram}
\newacronym{ecg}{ECG}{electrocardiogram}
\newacronym{fph}{FPH}{false positive per hour}
\newacronym{cnn}{CNN}{Convolutional Neural Network}
\newacronym{ban}{BAN}{Body Area Network}
\newacronym{roc}{ROC}{Receiver Operating Characteristic Curve}
\newacronym{auc}{AUC}{Area Under the Curve}
\newacronym{ecdf}{eCDF}{Empirical Cumulative Distribution Function}
\newacronym{deepchem}{DeepChEmM}{Deep Learning Channel Emulation Model}
\newacronym{svm}{SVM}{Support Vector Machine}
\newacronym{rbf}{RBF}{Radial Basis Function}
\newacronym{knn}{KNN}{$k$-Nearest Neighbors}
\newacronym{gbsm}{GBSM}{Geometry-Based Stochastic Model}
\newacronym{mlp}{MLP}{Multilayer Perceptron}
\newacronym{fc}{FC}{Fully Connected}
\newacronym{ae}{AE}{Autoencoder}
\newacronym{olla}{OLLA}{Outer Loop Link Adaptation}
\newacronym{rms}{RMS}{Root Mean Square}
\newacronym{ce}{CE}{Compute Engine}
\newacronym{tdl}{TDL}{tapped-delay-line}
\newacronym{bram}{BRAM}{block RAM}
\newacronym{rfnoc}{RFNoC}{RF Network-on-Chip}
\newacronym{ris}{RIS}{Reconfigurable Intelligent Surface}
\newacronym{dsp}{DSP}{Digital Signal Processing}
\newacronym{rtl}{RTL}{Register-Transfer Level}
\newacronym{hil}{HIL}{Hardware-in-the-loop}
\newacronym{oal}{OAL}{Open Air Link}
\newacronym{asic}{ASIC}{Application-Specific Integrated Circuit}
\newacronym{uram}{URAM}{UltraRAM}
\newacronym{ntn}{NTN}{Non-Terrestrial Network}
\newacronym{acf}{ACF}{Autocorrelation Function}
\newacronym{ks}{KS}{Kolmogorov--Smirnov}
\newacronym{psd}{PSD}{Power Spectral Density}

%% file: Figures/lchem-3ch.tex
\begin{tikzpicture}[
    x=1pt, y=-1pt,
    font=\fontsize{6}{7}\selectfont,
    block/.style={rectangle, rounded corners=1.7pt, line width=0.5pt,
      minimum width=20pt, minimum height=15pt, inner sep=0pt, align=center},
    uplink block/.style={block, draw=lchemDL2!80!black,
      fill=lchemDL2!14, text=lchemDL2!45!black},
    downlink block/.style={block, draw=lchemGainP5!80!black,
      fill=lchemGainP5!12, text=lchemGainP5!55!black},
    converter/.style={block, minimum height=15pt,
      draw=lchemDL0!80!black, fill=lchemDL0!14, text=lchemDL0!60!black},
    frontend/.style={block, minimum height=17pt,
      font=\fontsize{5.5}{6.5}\selectfont,
      draw=lchemCorrected!80!black, fill=lchemCorrected!10,
      text=lchemCorrected!80!black},
    group/.style={draw=lchemREF, line width=0.5pt,
      densely dashed, rounded corners=3pt},
    rf group/.style={draw=lchemREF!70, line width=0.5pt,
      dash pattern=on 4pt off 2pt on 0.5pt off 2pt, rounded corners=3pt},
    tip/.style={-{Latex[length=2.4pt, width=2.1pt]}},
    dl/.style={draw=lchemGainP5, line width=0.7pt, tip},
    ul/.style={draw=lchemDL2, line width=0.7pt, tip},
    crossing/.style={preaction={draw=white, line width=2.5pt, arrows=-}},
    analog/.style={draw=lchemREF, line width=0.55pt, tip},
    control/.style={draw=lchemDL1, densely dashed, line width=0.7pt, tip},
  ]
  \draw[group, draw=lchemGainBase] (0,0) rectangle (251,69);
  \node[text=lchemGainBase, anchor=south, inner sep=0pt]
    at (125.5,-2) {X410 FPGA / RFNoC Core};

  \foreach \x/\board in {0/0, 127/1} {
    \draw[group] (\x,82) rectangle (\x+124,157);
    \node[text=lchemGainBase] at (\x+62,151)
      {ZBX daughterboard \#\board};
  }
  \foreach \x/\rf/\id in {5/0/0, 68/1/1, 132/0/2, 195/1/3} {
    \draw[rf group] (\x,86) rectangle (\x+51,138);
    \node[converter] (dac\id) at (\x+12,95.5) {DAC};
    \node[converter] (adc\id) at (\x+39,95.5) {ADC};
    \node[frontend] (tx\id) at (\x+12,119) {RF\\Frontend};
    \node[frontend] (rx\id) at (\x+39,119) {RF\\Frontend};
    \draw[analog] (dac\id.south) -- (tx\id.north);
    \draw[analog] (rx\id.north) -- (adc\id.south);
    \node[text=lchemREF, font=\fontsize{5.5}{6.5}\selectfont]
      at (\x+12,133) {TX/RX 0};
    \node[text=lchemREF, font=\fontsize{5.5}{6.5}\selectfont]
      at (\x+39,133) {RX 1};
    \node[text=lchemREF] at (\x+25.5,142) {RF \rf};
  }

  \node[uplink block, minimum width=26pt, minimum height=11pt]
    (add1) at (33,26) {addsub1};
  \node[uplink block, minimum width=26pt, minimum height=11pt]
    (add0) at (70,10) {addsub0};
  \node[uplink block] (ulshift0) at (64,27) {Bit\\Shifter};
  \node[uplink block] (ulfir0) at (90,29) {FIR\\Filter};
  \node[uplink block] (ulshift1) at (122,27) {Bit\\Shifter};
  \node[uplink block] (ulfir1) at (150,27) {FIR\\Filter};
  \node[uplink block] (ulshift2) at (185,15) {Bit\\Shifter};
  \node[uplink block] (ulfir2) at (214,15) {FIR\\Filter};

  \node[downlink block, minimum width=19pt, minimum height=10pt]
    (split0) at (35,47) {split0};
  \node[downlink block, minimum width=20pt, minimum height=10pt]
    (split1) at (62,42) {split1};
  \node[downlink block] (dlfir0) at (61,57) {FIR\\Filter};
  \node[downlink block] (dlshift0) at (88,57) {Bit\\Shifter};
  \node[downlink block] (dlfir1) at (124,51) {FIR\\Filter};
  \node[downlink block] (dlshift1) at (151,51) {Bit\\Shifter};
  \node[downlink block] (dlfir2) at (181,39) {FIR\\Filter};
  \node[downlink block] (dlshift2) at (210,39) {Bit\\Shifter};

  \draw[dl] (adc0.north) -- (44,77) -- (19,77) -- (19,47)
    -- (split0.west);
  \node[anchor=west, inner sep=0pt, text=lchemGainP5] at (23,62) {DL};
  \draw[dl] (44.5,44.5) -- (48,44.5) -- (48,42) -- (split1.west);
  \draw[dl] (44.5,49.5) -- (46,49.5) -- (46,57) -- (dlfir0.west);
  \draw[dl] (dlfir0.east) -- (dlshift0.west);
  \draw[dl] (dlshift0.east) -- (102,57) -- (102,73)
    -- (80,73) -- (dac1.north);
  \draw[dl] (72,44.5) -- (100,44.5) -- (100,51) -- (dlfir1.west);
  \draw[dl] (dlfir1.east) -- (dlshift1.west);
  \draw[dl] (dlshift1.east) -- (162,51) -- (162,73)
    -- (144,73) -- (dac2.north);
  \draw[dl] (72,39.5) -- (161,39.5) -- (161,39) -- (dlfir2.west);
  \draw[dl] (dlfir2.east) -- (dlshift2.west);
  \draw[dl] (dlshift2.east) -- (224,39) -- (224,73)
    -- (207,73) -- (dac3.north);

  \draw[ul, crossing] (adc1.north) -- (107,29) -- (ulfir0.east);
  \draw[ul] ([yshift=2pt]ulfir0.west) -- (ulshift0.east);
  \draw[ul] (ulshift0.west) -- (46,27);
  \draw[ul, crossing] (adc2.north) -- (171,75) -- (166,75)
    -- (166,27) -- (ulfir1.east);
  \draw[ul] (ulfir1.west) -- (ulshift1.east);
  \draw[ul] (ulshift1.west) -- (102,27) -- (102,12.75) -- (83,12.75);
  \draw[ul, crossing] (adc3.north) -- (234,15) -- (ulfir2.east);
  \draw[ul] (ulfir2.west) -- (ulshift2.east);
  \draw[ul] (ulshift2.west) -- (129,15) -- (129,7.25) -- (83,7.25);
  \draw[ul] (add0.west) -- (50,10) -- (50,23.25) -- (46,23.25);
  \draw[ul] (add1.west) -- (17,26) -- (dac0.north);
  \node[anchor=east, inner sep=0pt, text=lchemDL2] at (13,53) {UL};
\end{tikzpicture}

%% file: Figures/fpga_dsp_heatmap.tex
\begin{tikzpicture}
  \begin{axis}[
    lchem axis,
    width=0.53\columnwidth,
    height=0.53\columnwidth,
    enlargelimits=false,
    view={0}{90},
    colormap name=lchemSeq,
    colorbar,
    colorbar style={width=0.025\columnwidth, tick align=outside,
      tick label style={font=\scriptsize},
      ytick={65,75,85,95,105},
      ylabel={DSP utilization (\%)},
      every axis label/.append style={font=\scriptsize}, ylabel near ticks},
    point meta min=65, point meta max=105,
    xtick={0,1,2,3}, xticklabels={4,8,16,32},
    ytick={0,1,2,3,4,5,6,7,8,9,10,11},
    yticklabels={1,2,3,4,5,6,7,8,9,10,11,12},
    xmin=-0.5, xmax=3.5, ymin=-0.5, ymax=11.5,
    xlabel={Complex taps per filter, $N_t$},
    ylabel={Directional filters, $N_\mathrm{ch}$},
    tick label style={font=\scriptsize},
    tick style={draw=none},
    grid=none,
  ]
    \addplot[matrix plot*, mesh/ordering=x varies, mesh/cols=4,
      point meta=explicit, draw=white, line width=0.3pt]
      table[col sep=comma, x=j, y=i, meta=util_pct]
      {data/fpga_dsp_heatmap_supported.csv};
    \node[text=black, font=\scriptsize] at (axis cs:0,0) {65.9};
    \node[text=black, font=\scriptsize] at (axis cs:1,0) {66.3};
    \node[text=black, font=\scriptsize] at (axis cs:2,0) {67.0};
    \node[text=black, font=\scriptsize] at (axis cs:3,0) {68.5};
    \node[text=black, font=\scriptsize] at (axis cs:0,1) {66.5};
    \node[text=black, font=\scriptsize] at (axis cs:1,1) {67.2};
    \node[text=black, font=\scriptsize] at (axis cs:2,1) {68.7};
    \node[text=black, font=\scriptsize] at (axis cs:3,1) {71.7};
    \node[text=black, font=\scriptsize] at (axis cs:0,2) {67.0};
    \node[text=black, font=\scriptsize] at (axis cs:1,2) {68.1};
    \node[text=black, font=\scriptsize] at (axis cs:2,2) {70.4};
    \node[text=black, font=\scriptsize] at (axis cs:3,2) {74.9};
    \node[text=black, font=\scriptsize] at (axis cs:0,3) {67.6};
    \node[text=black, font=\scriptsize] at (axis cs:1,3) {69.1};
    \node[text=black, font=\scriptsize] at (axis cs:2,3) {72.1};
    \node[text=black, font=\scriptsize] at (axis cs:3,3) {78.1};
    \node[text=black, font=\scriptsize] at (axis cs:0,4) {68.1};
    \node[text=black, font=\scriptsize] at (axis cs:1,4) {70.0};
    \node[text=black, font=\scriptsize] at (axis cs:2,4) {73.8};
    \node[text=black, font=\scriptsize] at (axis cs:3,4) {81.2};
    \node[text=black, font=\scriptsize] at (axis cs:0,5) {68.7};
    \node[text=black, font=\scriptsize] at (axis cs:1,5) {71.0};
    \node[text=black, font=\scriptsize] at (axis cs:2,5) {\bfseries 75.4};
    \node[text=black, font=\scriptsize] at (axis cs:3,5) {84.4};
    \node[text=black, font=\scriptsize] at (axis cs:0,6) {69.3};
    \node[text=black, font=\scriptsize] at (axis cs:1,6) {71.9};
    \node[text=black, font=\scriptsize] at (axis cs:2,6) {77.1};
    \node[text=black, font=\scriptsize] at (axis cs:3,6) {87.6};
    \node[text=black, font=\scriptsize] at (axis cs:0,7) {69.8};
    \node[text=black, font=\scriptsize] at (axis cs:1,7) {72.8};
    \node[text=black, font=\scriptsize] at (axis cs:2,7) {78.8};
    \node[text=black, font=\scriptsize] at (axis cs:3,7) {90.8};
    \node[text=black, font=\scriptsize] at (axis cs:0,8) {70.4};
    \node[text=black, font=\scriptsize] at (axis cs:1,8) {73.8};
    \node[text=black, font=\scriptsize] at (axis cs:2,8) {80.5};
    \node[text=black, font=\scriptsize] at (axis cs:3,8) {94.0};
    \node[text=black, font=\scriptsize] at (axis cs:0,9) {71.0};
    \node[text=black, font=\scriptsize] at (axis cs:1,9) {74.7};
    \node[text=black, font=\scriptsize] at (axis cs:2,9) {82.2};
    \node[text=white, font=\scriptsize] at (axis cs:3,9) {97.2};
    \node[text=black, font=\scriptsize] at (axis cs:0,10) {71.5};
    \node[text=black, font=\scriptsize] at (axis cs:1,10) {75.6};
    \node[text=black, font=\scriptsize] at (axis cs:2,10) {83.9};
    \node[text=white, font=\scriptsize] at (axis cs:3,10) {100.4};
    \node[text=black, font=\scriptsize] at (axis cs:0,11) {72.1};
    \node[text=black, font=\scriptsize] at (axis cs:1,11) {76.6};
    \node[text=black, font=\scriptsize] at (axis cs:2,11) {85.6};
    \node[text=white, font=\scriptsize] at (axis cs:3,11) {103.5};

    \draw[black, densely dashed, line width=1pt]
      (axis cs:3.5,3.5) -- (axis cs:2.5,3.5) --
      (axis cs:2.5,7.5) -- (axis cs:1.5,7.5) -- (axis cs:1.5,11.5);
    \draw[black, line width=1.2pt]
      (axis cs:1.56,4.56) rectangle (axis cs:2.44,5.44);
  \end{axis}
\end{tikzpicture}

%% file: Figures/lchem_validation.tex
\begin{tikzpicture}[
    x=1pt, y=-1pt,
    font=\scriptsize,
    blue box/.style={draw=lchemDL0!80!black, fill=lchemDL0!14, line width=0.5pt, rounded corners=2pt},
    red box/.style={draw=lchemDL1!80!black, fill=lchemDL1!12, line width=0.5pt, rounded corners=2pt},
    green box/.style={draw=lchemDL2!80!black, fill=lchemDL2!14, line width=0.5pt, rounded corners=2pt},
    orange box/.style={draw=lchemGainP5!80!black, fill=lchemGainP5!12, line width=0.5pt, rounded corners=2pt},
    container/.style={line width=0.6pt, dash dot, rounded corners=3pt},
    rf/.style={draw=lchemDL0, line width=1pt},
    txin/.style={draw=lchemGainP5, line width=1pt},
    lan/.style={draw=lchemDL1, line width=0.8pt, dash pattern=on 3pt off 1.5pt},
    emulated/.style={draw=lchemDL2, line width=1pt, rounded corners=2pt},
    chan/.style={draw=lchemDL2, line width=0.6pt, densely dashed},
    tip/.style={-{Latex[length=3.5pt, width=3pt]}},
    on line/.style={fill=white, inner sep=1pt},
  ]
  \draw[blue box] (0,0) rectangle (72,20);
  \node[align=center] at (36,10) {USRP X410\\Transmitter \textbar\ GLFSR};
  \draw[blue box] (0,37) rectangle (72,57);
  \node[align=center] at (36,47) {RF Splitter\\Passive 1:2};
  \draw[red box] (112,0) rectangle (162,20);
  \node[align=center] at (137,10) {Experiment PC\\Python / SSH};

  \draw[rf, tip] (20,20) -- (20,37);
  \node[on line] at (20,27.5) {RF};

  \draw[container] (182,20) rectangle (251.5,145);
  \node[align=center] at (216.75,30) {Keysight MXR608A\\Oscilloscope};
  \draw[blue box] (184.5,40) rectangle (249,54);
  \node at (216.75,47) {Ch1---Reference};
  \foreach \y/\n/\rx in {72/2/1, 96/3/2, 120/4/3} {
    \draw[green box] (184.5,\y-7) rectangle (249,\y+7);
    \node at (216.75,\y) {Ch\n---Emulated RX\rx};
  }

  \draw[rf, tip] (72,47) -- (184.5,47);
  \node[on line] at (100,47) {RF (reference)};

  \draw[lan, tip] (112,10) -- (72,10);
  \node at (92,5) {LAN};
  \node at (92,15.5) {(TX control)};
  \draw[lan, tip] (137,20) -- (137,62);
  \node[align=right, anchor=east] at (134.5,31) {LAN\\(channel control)};
  \draw[lan, tip] (162,10) -- (216.75,10) -- (216.75,20);
  \node[anchor=south] at (205,9.5) {LAN (waveform capture)};

  \draw[container] (0,62) rectangle (165,145);
  \node[align=center, anchor=north] at (79,63.5) {USRP X410\\Channel Emulator};
  \draw[orange box] (2,108) rectangle (34,122);
  \node at (18,115) {TX Input};
  \foreach \x/\n in {43/1, 84/2, 125/3} {
    \draw[green box] (\x,108) rectangle (\x+38,122);
    \node at (\x+19,115) {RX\n\ Output};
  }
  \draw[txin, tip] (20,57) -- (20,108);
  \node[on line] at (20,72) {RF};

  \draw[chan, tip] (34,115) -- (43,115);
  \draw[chan, tip] (28,108) -- (28,99) -- (62,99)  -- (62,108);
  \draw[chan, tip] (12,108) -- (12,91) -- (103,91) -- (103,108);
  \draw[chan, tip] (6,108)  -- (6,83)  -- (144,83) -- (144,108);
  \node[on line, text=lchemDL2!80!black] at (45,99)  {$*\,h_1(t)$};
  \node[on line, text=lchemDL2!80!black] at (82,91)  {$*\,h_2(t)$};
  \node[on line, text=lchemDL2!80!black] at (118,83) {$*\,h_3(t)$};

  \draw[emulated, tip] (62,122)  -- (62,140)  -- (168,140) -- (168,72)  -- (184.5,72);
  \draw[emulated, tip] (103,122) -- (103,134) -- (172,134) -- (172,96)  -- (184.5,96);
  \draw[emulated, tip] (144,122) -- (144,128) -- (176,128) -- (176,120) -- (184.5,120);

\end{tikzpicture}

%% file: Figures/delay_error.tex
\begin{tikzpicture}
  \begin{axis}[
    lchem panel,
    xlabel={Configured delay ($\mu$s)},
    ylabel={Delay error (ps)},
    xmin=0, xmax=17, ymin=-60, ymax=60,
  ]
    \addplot[lchem ideal, no marks, forget plot]
      coordinates {(0,0) (17,0)};

    \addplot[lchem measured,
             error bars/.cd, y dir=both, y explicit]
      table[col sep=comma, x=expected_us, y expr=\thisrow{error_ns}*1000, y error expr=\thisrow{std_ns}*1000]
      {data/delay_error.csv};
  \end{axis}
\end{tikzpicture}

%% file: Figures/experimental_setup.tex
\begin{tikzpicture}[
    x=0.74pt, y=-0.74pt,
    font=\scriptsize,
    box/.style={draw=lchemDL0!80!black, fill=lchemDL0!14, line width=0.5pt, rounded corners=2pt},
    att/.style={draw=lchemGainP5!75!black, fill=lchemGainP5!12, line width=0.5pt, rounded corners=2pt},
    port/.style={draw=lchemCorrected!80!black, fill=lchemCorrected!10, line width=0.5pt, rounded corners=1.5pt},
    unused/.style={draw=lchemCorrected!30, fill=lchemCorrected!3, line width=0.5pt, rounded corners=1.5pt},
    group/.style={draw=lchemREF, line width=0.5pt, dash dot, rounded corners=2pt},
    dl/.style={draw=lchemGainP5, line width=1pt},
    ul/.style={draw=lchemDL2, line width=1pt},
    both/.style={draw=lchemCorrected!80!black, line width=1pt},
    tip/.style={-{Latex[length=3.5pt, width=3pt]}},
  ]
  \draw[line width=0.6pt, dash dot, rounded corners=3pt] (0.8,0.4) rectangle (337.7,70.0);
  \node[align=center] at (289.5,15.3) {USRP X410\\Channel Emulator};

  \foreach \x/\rf in {7.2/0, 87.0/1, 173.9/0, 253.6/1} {
    \draw[group] (\x,31.0) rectangle ++(77.6,34.5);
    \node[text=lchemREF] at (\x+38.8,59.5) {RF \rf};
  }
  \foreach \x/\name in {9.7/{TX/RX\,0}, 47.2/{RX\,1}, 176.4/{TX/RX\,0}, 213.9/{RX\,1}} {
    \draw[port] (\x,34.5) rectangle ++(35.6,18.5);
    \node[text=lchemCorrected!80!black] at (\x+17.8,43.75) {\name};
  }
  \foreach \x/\name in {89.5/{TX/RX\,0}, 127.0/{RX\,1}, 256.1/{TX/RX\,0}, 293.6/{RX\,1}} {
    \draw[unused] (\x,34.5) rectangle ++(35.6,18.5);
    \node[text=lchemCorrected!40] at (\x+17.8,43.75) {\name};
  }

  \draw[ul, dashed, tip] (231.7,34.5) -- (231.7,10.4) -- (27.5,10.4) -- (27.5,34.5);
  \draw[dl, dashed, tip] (65.0,34.5) -- (65.0,21.7) -- (194.2,21.7) -- (194.2,34.5);
  \node[fill=white, inner sep=1pt, text=lchemDL2] at (93.3,10.4) {$*\,h_{\mathrm{UL}}(t)$};
  \node[fill=white, inner sep=1pt, text=lchemGainP5] at (150.3,21.7) {$*\,h_{\mathrm{DL}}(t)$};

  \draw[draw=lchemGainBase, fill=lchemGainBase!7, line width=0.8pt, rounded corners=3pt]
    (0.8,98.5) rectangle (168.2,144.5);
  \node[text=lchemGainBase] at (130.3,107.5) {RF Shield Box};
  \draw[box] (13.2,113.5) rectangle (78.8,140.5);
  \node[align=center] at (46.0,127.0) {RF Splitter\\Passive 1:2};
  \draw[box] (94.4,113.5) rectangle (161.4,140.5);
  \node[align=center] at (127.9,127.0) {Radio Unit\\Foxconn};
  \draw[box] (88.7,154.5) rectangle (167.1,184.5);
  \node[align=center] at (127.9,169.5) {gNB\\OpenAirInterface};
  \draw[att] (43.65,78.0) rectangle (86.35,92.2);
  \node[text=lchemGainP5!55!black] at (65.0,85.1) {20\,dB Att.};

  \draw[both] (78.8,127.0) -- (94.4,127.0);
  \draw[draw=lchemTheory, line width=0.8pt, {Latex[length=3pt, width=2.6pt]}-{Latex[length=3pt, width=2.6pt]}]
    (127.9,140.5) -- (127.9,154.5);
  \node[text=lchemTheory, anchor=west] at (131.0,147.5) {FH};

  \draw[ul, tip] (27.5,53.0) -- (27.5,113.5);
  \draw[dl, tip] (65.0,113.5) -- (65.0,92.2);
  \draw[dl, tip] (65.0,78.0) -- (65.0,53.0);

  \draw[att] (172.8,78.0) rectangle (215.5,92.2);
  \node[text=lchemGainP5!55!black] at (194.2,85.1) {20\,dB Att.};
  \draw[box] (212.65,122.7) circle [radius=23.5];
  \node at (212.65,122.7) {Circulator};
  \draw[box] (180.6,154.5) rectangle (244.7,184.5);
  \node[align=center] at (212.65,169.5) {RF Splitter\\Passive 1:4};
  \draw[box] (259.0,154.5) rectangle (337.3,184.5);
  \node[align=center] at (298.2,169.5) {User Equipment\\Sierra Wireless};

  \draw[dl, tip] (194.2,53.0) -- (194.2,78.0);
  \draw[dl, tip] (194.2,92.2) -- (194.2,108.15);   
  \draw[ul, tip] (231.7,108.94) -- (231.7,53.0);   
  \draw[both] (212.65,146.2) -- (212.65,154.5);    
  \foreach \y in {158.25, 165.75, 173.25, 180.75}
    \draw[both] (244.7,\y) -- (259.0,\y);
  \node[text=lchemGainBase] at (185.8,102.5) {2};
  \node[text=lchemGainBase] at (240.2,102.5) {1};
  \node[text=lchemGainBase] at (220.8,150.5) {3};

  \draw[dl, tip] (261.5,102.0) -- (281.2,102.0);
  \node[anchor=west] at (288.0,102.0) {DL};
  \draw[ul, tip] (261.5,114.0) -- (281.2,114.0);
  \node[anchor=west] at (288.0,114.0) {UL};
  \draw[both] (261.5,126.0) -- (281.2,126.0);
  \node[anchor=west] at (288.0,126.0) {DL + UL};
  \draw[unused] (261.8,133.5) rectangle (281.2,142.5);
  \node[anchor=west] at (288.0,138.0) {Unused};
\end{tikzpicture}

%% file: Figures/ran_gain_snr_rsrp.tex
%
%
%
\begin{tikzpicture}
  \pgfplotsset{
    gainaxis/.style={
      lchem panel,
      xmin=0, xmax=55, ymin=-11, ymax=55,
      xtick={0,10,20,30,40,50},
      enlarge x limits=false,
    },
    gainline/.style={line width=0.8pt, no marks},
  }

  \begin{axis}[
    gainaxis,
    name=mainax,
    xlabel={Emulated path loss (dB)},
    ylabel={DL SNR (dB)},
    ytick={0,20,40},
    unbounded coords=discard,
  ]
    \addplot[gainline, color=lchemGainM10]  table[col sep=comma, x=pl_db, y=snr_m10]  {data/ran_gain_snr_rsrp.csv};
    \addplot[gainline, color=lchemGainM5]   table[col sep=comma, x=pl_db, y=snr_m5]   {data/ran_gain_snr_rsrp.csv};
    \addplot[gainline, color=lchemGainBase] table[col sep=comma, x=pl_db, y=snr_base] {data/ran_gain_snr_rsrp.csv};
    \addplot[gainline, color=lchemGainP5]   table[col sep=comma, x=pl_db, y=snr_p5]   {data/ran_gain_snr_rsrp.csv};
    \addplot[gainline, color=lchemGainP10]  table[col sep=comma, x=pl_db, y=snr_p10]  {data/ran_gain_snr_rsrp.csv};

    \addplot[gainline, densely dotted, color=lchemGainM10,  forget plot] table[col sep=comma, x=pl_db, y=rsrpa_m10]  {data/ran_gain_snr_rsrp.csv};
    \addplot[gainline, densely dotted, color=lchemGainM5,   forget plot] table[col sep=comma, x=pl_db, y=rsrpa_m5]   {data/ran_gain_snr_rsrp.csv};
    \addplot[gainline, densely dotted, color=lchemGainBase, forget plot] table[col sep=comma, x=pl_db, y=rsrpa_base] {data/ran_gain_snr_rsrp.csv};
    \addplot[gainline, densely dotted, color=lchemGainP5,   forget plot] table[col sep=comma, x=pl_db, y=rsrpa_p5]   {data/ran_gain_snr_rsrp.csv};
    \addplot[gainline, densely dotted, color=lchemGainP10,  forget plot] table[col sep=comma, x=pl_db, y=rsrpa_p10]  {data/ran_gain_snr_rsrp.csv};

  \end{axis}

  \begin{axis}[
    gainaxis,
    at=(mainax.south west), anchor=south west,
    axis x line=none,
    axis y line*=right,
    tick pos=right,
    ylabel near ticks,
    y tick label style={anchor=west},
    grid=none,
    ylabel={RSRP (dBm)},
    ymin=-121.79, ymax=-55.79,          
    ytick={-120,-100,-80,-60},
  ]
  \end{axis}
\end{tikzpicture}

%% file: Figures/ran_gain_ul_uetx_pgf.tex
%
%
%
\begin{tikzpicture}
  \begin{axis}[
    lchem panel,
    xlabel={Emulated path loss (dB)},
    ylabel={UE Tx power (dBm)},
    xmin=0, xmax=45, ymin=-24, ymax=17,
    xtick={0,10,20,30,40},
    ytick={-20,-10,0,10},
    enlarge x limits=false,
    unbounded coords=discard,
  ]
    \addplot[lchem ideal, no marks, forget plot]
      coordinates {(0,7) (45,7)};
    \node[font=\scriptsize, anchor=south east] at (axis cs:44.5,7.2)
      {$P_\mathrm{CMAX}$};

    \addplot[line width=0.8pt, no marks, color=lchemGainM10]
      table[col sep=comma, x=pl_db, y=uetx_m10]  {data/ran_gain_ul_uetx.csv};
    \addplot[line width=0.8pt, no marks, color=lchemGainM5]
      table[col sep=comma, x=pl_db, y=uetx_m5]   {data/ran_gain_ul_uetx.csv};
    \addplot[line width=0.8pt, no marks, color=lchemGainBase]
      table[col sep=comma, x=pl_db, y=uetx_base] {data/ran_gain_ul_uetx.csv};
    \addplot[line width=0.8pt, no marks, color=lchemGainP5]
      table[col sep=comma, x=pl_db, y=uetx_p5]   {data/ran_gain_ul_uetx.csv};
    \addplot[line width=0.8pt, no marks, color=lchemGainP10]
      table[col sep=comma, x=pl_db, y=uetx_p10]  {data/ran_gain_ul_uetx.csv};
  \end{axis}
\end{tikzpicture}

%% file: Figures/ran_gain_legend.tex
\begingroup
\newcommand{\lchemGainSample}[1]{%
  \tikz[baseline=-0.5ex]{\draw[#1] (0,0) -- (0.50cm,0);}%
}
\begin{tikzpicture}
  \node[draw=lchemGrid, fill=white, font=\scriptsize,
    inner xsep=2pt, inner ysep=1.5pt] {%
    \begin{tabular}{@{}c@{}}
      \begin{tabular}{@{}l@{\hspace{6pt}}l@{\hspace{6pt}}l@{\hspace{6pt}}l@{\hspace{6pt}}l@{}}
        \lchemGainSample{line width=0.8pt, color=lchemGainM10}\,$-10$\,dB &
        \lchemGainSample{line width=0.8pt, color=lchemGainM5}\,$-5$\,dB &
        \lchemGainSample{line width=0.8pt, color=lchemGainBase}\,blueprint &
        \lchemGainSample{line width=0.8pt, color=lchemGainP5}\,$+5$\,dB &
        \lchemGainSample{line width=0.8pt, color=lchemGainP10}\,$+10$\,dB
      \end{tabular}\\[2pt]
      \lchemGainSample{solid, line width=1pt, color=black}\,SNR
      \hspace{12pt}
      \lchemGainSample{densely dotted, line width=1pt, color=black}\,RSRP
    \end{tabular}%
  };
\end{tikzpicture}
\endgroup